\documentclass[preprint]{aastex61}
\usepackage{graphicx,natbib} 
\shorttitle{CALIBRATION OF STAR FORMATION RATE INDICATORS}
\shortauthors{Brown et al.}

\date{\today}

\begin{document}

\title{CALIBRATION OF ULTRAVIOLET, MID-INFRARED and RADIO STAR FORMATION RATE INDICATORS}

\correspondingauthor{Michael J. I. Brown}
\email{Michael.Brown@monash.edu}

\author[0000-0002-1207-9137]{Michael J. I. Brown}
\affil{School of Physics and Astronomy, Monash University, Clayton, Victoria 3800, Australia}
\affil{Monash Centre for Astrophysics, Monash University, Clayton, Victoria, 3800, Australia}

\author{John Moustakas}
\affil{Department of Physics and Astronomy, Siena College, 515 Loudon Road, Loudonville, NY 12211, USA}

\author{Robert C. Kennicutt}
\affil{Institute of Astronomy, University of Cambridge, Cambridge, CB3 0HA, United Kingdom}

\author{Nicolas J. Bonne}
\affil{Institute for Cosmology and Gravitation, Dennis Sciama Building, University of Portsmouth, Burnaby Road, Portsmouth PO1 3FX, United Kingdom}

\author{Huib T. Intema}
\affil{Leiden Observatory, Leiden University, P.O. Box 9513, 2300 RA Leiden, The Netherlands}

\author{Francesco de Gasperin}
\affil{Leiden Observatory, Leiden University, P.O. Box 9513, 2300 RA Leiden, The Netherlands}

\author{Mederic Boquien}
\affil{Universidad de Antofagasta, Unidad de Astronomia, Avenida Angamos 601, 02800 Antofagasta, Chile}
\affil{Institute of Astronomy, University of Cambridge, Cambridge, CB3 0HA, United Kingdom}

\author{T. H. Jarrett}
\affil{Astrophysics, Cosmology and Gravity Centre (ACGC), Astronomy Department, University of Cape Town, Private Bag X3, Rondebosch 7701, South Africa}

\author{Michelle E. Cluver}
\affil{University of the Western Cape, Robert Sobukwe Road, Bellville 7535, South Africa}

\author{J.-D. T. Smith}
\affil{Department of Physics and Astronomy, University of Toledo, Ritter Obs., MS \#113, Toledo, OH 43606, USA}

\author{Elisabete da Cunha}
\affil{Research School of Astronomy and Astrophysics, Australian National University, Canberra, ACT 2611, Australia}
\affil{Centre for Astrophysics and Supercomputing, Swinburne University of Technology, Hawthorn, Victoria 3122, Australia}

\author{Masatoshi Imanishi}
\affil{Subaru Telescope, 650 North A'ohoku Place, Hilo, HI 96720, USA}
\affil{Department of Astronomy, School of Science, Graduate University for Advanced Studies (SOKENDAI), Mitaka, Tokyo 181-8588, Japan}
\affil{National Astronomical Observatory of Japan, 2-21-1 Osawa, Mitaka, Tokyo 181-8588, Japan}

\author{Lee Armus}
\affil{Spitzer Science Center, California Institute of Technology, Pasadena, CA, USA}

\author{Bernhard R. Brandl}
\affil{Leiden Observatory, Leiden University, P.O. Box 9513, 2300 RA Leiden, The Netherlands}

\author{J. E. G. Peek}
\affil{Space Telescope Science Institute, 3700 San Martin Dr., Baltimore, MD 21218, USA}
\affil{Department of Astronomy, Columbia University, New York, NY, USA}

\begin{abstract}
We present calibrations for star formation rate indicators in the ultraviolet, mid-infrared and radio continuum bands, including one of the first direct calibrations of $150~{\rm MHz}$ as a star formation rate indicator. Our calibrations utilize 66 nearby star forming galaxies with Balmer decrement corrected ${\rm H\alpha}$ luminosities, which span 5 orders of magnitude in star formation rate and have absolute magnitudes of $-24<M_r<-12$. Most of our photometry and spectrophotometry is measured from the same region of each galaxy, and our spectrophotometry has been validated with SDSS photometry, so our random and systematic errors are small relative to the intrinsic scatter seen in star formation rate indicator calibrations. We find WISE $W4$ ($22.8~{\rm \mu m}$), {\it Spitzer} $24~{\rm \mu m}$ and $1.4~{\rm GHz}$ have tight correlations with Balmer decrement corrected H$\alpha$ luminosity, with scatter of only 0.2~dex. Our calibrations are comparable to those from the prior literature for $L^*$ galaxies, but for dwarf galaxies our calibrations can give star formation rates that are far greater than those derived from much of the prior literature.  
\end{abstract}
\keywords{dust, extinction --- galaxies: general --- galaxies: evolution --- galaxies: photometry --- stars: formation ---  techniques: spectroscopic} 


\section{Introduction}

Galaxies increase their stellar masses via star formation and mergers, and thus measurements of galaxy star formation rates (SFRs) are critical for many observational studies of galaxy evolution. In principle, very accurate star formation rates are provided by ultraviolet and hydrogen recombination line luminosities, which directly trace the population of short-lived very massive stars \citep[][and references therein]{ken12}. In practice, measured ultraviolet luminosities are sensitive to dust attenuation and accurate spectrophotometry is often unavailable or limited to the cores of galaxies. For example, the vast majority of galaxies in deep optical, mid-infrared and radio continuum surveys do not have spectroscopic redshifts, and this will remain true for the foreseeable future \citep[e.g.,][]{des05,pap06,nor11}.

As a consequence of the limitations of spectroscopy and ultraviolet imaging, a number of SFR indicators have been utilized at mid-infrared, far-infrared and radio wavelengths. For a detailed discussion of these SFR indicators and their calibration, we refer the reader to \citet{ken09}, \citet{ken12} and references therein. The integrated far-infrared emission is (comparatively) straightforward to understand, as it results from dust heated primarily by ultraviolet and optical photons from massive stars. However, at specific wavelengths the emission has a non-trivial relationship with SFR. For example, measurements with the WISE $W3$ ($12~{\rm \mu m}$) band can include contributions from thermal emission from dust, a deep silicate absorption feature and emission attributed to polycyclic aromatic hydrocarbons (PAHs), which has a metallicity dependence \citep[e.g.,][]{hou04b,eng05,jac06,dra07,smi07,eng08}. Furthermore, the thermal emission from dust can result from star formation, active galactic nuclei (AGN) and old stellar populations \citep[e.g.,][]{wal87,ben10,boq11}. It is possible to model the relationship between observed galaxy luminosities and star formation rates via detailed galaxy SED modeling \citep[e.g.,][]{dac08,boq16,dav16,lej16}, but a more common approach is to empirically calibrate SFR indicators using hydrogen recombination line luminosities with corrections for dust attenuation.

Although empirical calibrations of SFR indicators are far simpler than SED modeling, they are not completely free from modeling and the resulting model dependent assumptions. The relationship between ${\rm H\alpha}$ luminosity and SFR depends on the adopted stellar initial mass function (IMF), which may not be universal \citep[e.g.,][]{van10}, and the recent star formation history \citep[e.g.][]{wei12,das14}. Dust obscuration is often modeled using a dusty screen rather than more complex (and realistic, yet uncertain) dust geometries, and the Balmer decrement measurements of dust obscuration typically adopt a set of conditions for the interstellar gas that cannot apply throughout individual galaxies, let alone throughout entire galaxy populations \citep[e.g.,][ and references therein]{cal94,boq12}. Measurements of weak nebular emission lines in galaxy spectra rely on subtracting the stellar continuum, which requires modeling of star formation histories and stellar populations (including details such as metallicity). Relationships between SFR indicator and hydrogen recombination line luminosities are frequently modeled with linear relationships or power-laws, without clear physical motivation (although good fits can be achieved). That said, as discussed by \citet{ken09}, such simplified (and transparent) modeling can still produce reliable calibrations for SFR indicators consistent with more complicated modeling of galaxy SEDs. 

The empirical calibrations of SFR indicators are critically reliant on the accuracy of measurements of hydrogen recombination line fluxes, dust attenuation corrections and photometry, all of which present challenges. Achieving spectrophotometric accuracies better than 10\% is non-trivial and spectroscopy is often limited to galaxy cores (e.g., fiber-fed and slit spectroscopy), requiring aperture corrections to measure hydrogen recombination line fluxes for entire galaxies \citep[e.g.,][]{hop03, brou11}. Matching catalogues of emission line fluxes and catalogues of broadband photometry can be performed relatively quickly, but ideally spectra and photometry should be extracted from the same regions of individual galaxies (thus mitigating difficulties with aperture corrections). Reliable emission line fluxes require accurate subtraction of the continuum and absorption lines from stellar populations \citep[e.g.,][]{tre04,mou06}, and Balmer decrement corrections of dust attenuation require high signal-to-noise measurements of emission lines. Photometric zero-point errors, effective wavelength errors and other systematic errors (e.g., scattered light in the {\it Spitzer} IRAC detector) can hamper the calibration of star formation rate indicators. For example, in \citet{bro14} we identified an effective wavelength error in the WISE $W4$ filter curve, which results in the $22~{\rm \mu m}$ flux densities of luminous infrared galaxies (LIRGs) being overestimated by up to 30\%. 

Sample selection inevitably plays a role in SFR indicator calibrations. Magnitude limited samples are dominated by $\sim L^*$ galaxies that fall on the SFR - mass relation \citep[i.e., the ``star forming main sequence,''][]{noe07}, and have relatively few low luminosity dwarf galaxies and LIRGs. Many galaxy samples have minimum redshift, maximum size (e.g., for integral field or fiber fed spectroscopy) and maximum flux limits (e.g., to prevent cross-talk in multi-object spectroscopy), which effectively places limits on galaxy stellar masses and SFRs. For example, the \citet{clu14} calibration of the WISE $W3$ and $W4$ bands uses galaxies with SFRs greater than $10^{-1}~{\rm M_\odot~yr^{-1}}$. Consequently, a number of the SFR calibrations from the literature use samples with ${\rm H\alpha}$ luminosities that span less than three orders of magnitude \citep{wu05,lee13,clu14,cat15}, and extrapolations of such empirical calibrations obviously carry risks. 

SFR indicator calibrations have been extended to low SFRs using individual H~II regions, but the relationship between SFR indicator luminosity and SFR of H~II regions in $\sim L^*$ galaxies differs from that of dwarf galaxies \citep[e.g.,][]{cal07,rel07,ken09}. Prior to the widespread availability of {\it Spitzer} and WISE mid-infrared archival imaging, (IRAS) Infrared Astronomical Satellite photometry was used for mid-infrared SFR calibrations, which excludes low luminosity galaxies and potentially introduces errors when IRAS fluxes are used as proxies for {\it Spitzer} and WISE fluxes \citep{ken09}. Of course these issues are well-known to the relevant authors, who were generally using the best available data at the time of publication.

In this paper we present SFR calibrations for the GALEX $FUV$, {\it Spitzer} mid-infrared bands, WISE mid-infrared bands and radio continuum. Our focus is on monochromatic SFR indicators, in part due to the data we currently have available and in part because such calibrations will be readily usable by new deep wide-field surveys \citep[e.g.,][]{nor11,wil16}. The calibrations utilize the photometry and spectral energy distributions (SEDs) of \citet{bro14}, and new photometry of galaxies with distances of $\lesssim 10~{\rm Mpc}$. The bulk of the photometry and spectrophotometry is accurate to 10\%, and for most wavelengths our photometry and spectra are extracted from the same region of each galaxy, minimizing the impact of aperture corrections. Our galaxy sample spans $-24<M_r<-12$ and $-0.3<u-r<2.3$ (AB), and includes LIRGs and blue compact dwarfs, as well as regular $\sim L^*$ spiral galaxies. Balmer decrement corrected ${\rm H\alpha}$ luminosities, and thus star formation rates, span almost five orders of magnitude. We thus expect our SFR indicator calibrations to be applicable to a broader range of galaxies than many of the calibrations from the prior literature. 

The structure of this paper is as follows. Section~\ref{sec:data} presents an overview of the archival imaging, photometry and spectroscopy used in our study. In Section~\ref{sec:emflux} we discuss our new emission line flux measurements, which are critical for sample selection and Balmer decrement H$\alpha$ luminosity measurements. In Section~\ref{sec:sample} we describe the selection of the star forming galaxy sample and the basic observable properties of this sample (e.g., absolute magnitudes, colors). The calibration of SFR indicators is discussed in Section~\ref{sec:calibration} and our principal conclusions are summarized in Section~\ref{sec:summary}. Throughout this paper we use AB magnitudes and adopt a bolometric luminosity $3.827\times 10^{33}~{\rm erg~s^{-1}}$ for the Sun. To simplify comparison with the prior literature, broadband luminosities are $\nu L_\nu$ with units of ${\rm erg~s^{-1}}$, while radio powers are presented in units of ${\rm W~Hz^{-1}}$. 

\section{Data}
\label{sec:data}

Our parent sample is star-forming galaxies with optical drift-scan spectrophotometry from \citet{mou06} and \citet{mou10} that also have Sloan Digital Sky Survey III optical imaging \citep[SDSS III;][]{sdss3}. The extraction apertures for the optical spectrophotometry vary in size between $20^{\prime\prime} \times 20^{\prime\prime}$ and $\sim 15^{\prime} \times 3^{\prime}$, and thus the spectra include much of the relevant galaxy light. We presented the ultraviolet to mid-infrared photometry and SEDs for many of these galaxies in \citet{bro14}. For the galaxies that weren't previously presented in \citet{bro14}, the data sources and methods are effectively identical to those of \citet{bro14}.

All of the galaxies in the sample have imaging at ultraviolet, optical, near-infrared and mid-infrared wavelengths, taken from the Galaxy Evolution Explorer \citep[GALEX;][]{mor07},  {\it Swift} UV/optical monitor telescope \citep[UVOT;][]{rom05}, Sloan Digital Sky Survey III \citep[SDSS III;][]{sdss3}, Two Micron All Sky Survey \citep[2MASS;][]{skr06}, {\it Spitzer} Space Telescope \citep{faz04,rie04} and/or Wide-field Infrared Space Explorer \citep[WISE;][]{wri10}. Absolute photometric calibration for these imaging surveys is typically on the order of a few percent for stellar sources \citep{skr06,pad08,wri10,boh11,boh14}, although larger photometric calibration errors may be present in the UV \citep[GALEX calibration issues are discussed in detail by][]{cam14} and for extended source photometry \citep[e.g.,][]{jar11}. Foreground dust extinction was modeled using the Planck dust extinction maps \citep{pla11,pla13} and the \cite{fit99} extinction curve, with the modification to the UV attenuation proposed by \citet{pee2013b}. However, it should be noted that for the bulk of the galaxies in our sample the foreground dust extinction is less than $E(B-V)=0.05$.  


Matched aperture photometry was measured in all bands shortward of $30~{\rm \mu m}$ using the same rectangular aperture that was used for the optical drift-scan spectrophotometry. The methods used to measure the aperture photometry are largely identical to those of \cite{bro14}, including coincidence loss corrections for {\it Swift} photometry and scattered light corrections for {\it Spitzer} IRAC photometry. However, unlike \cite{bro14}, we corrected for the difference between between the in-orbit and laboratory measured WISE $W4$ effective wavelengths, using the method of \citet{bro14b}. Uncertainties were determined by measuring aperture photometry at positions offset from the galaxy position and then measuring the range that encompassed 68\% of the data. For most galaxies and bands the uncertainties are less than $0.1~{\rm mag}$, and for the SFR calibrations we exclude photometry if the uncertainties are greater than $0.2~{\rm mag}$.

\begin{figure}
\plotone{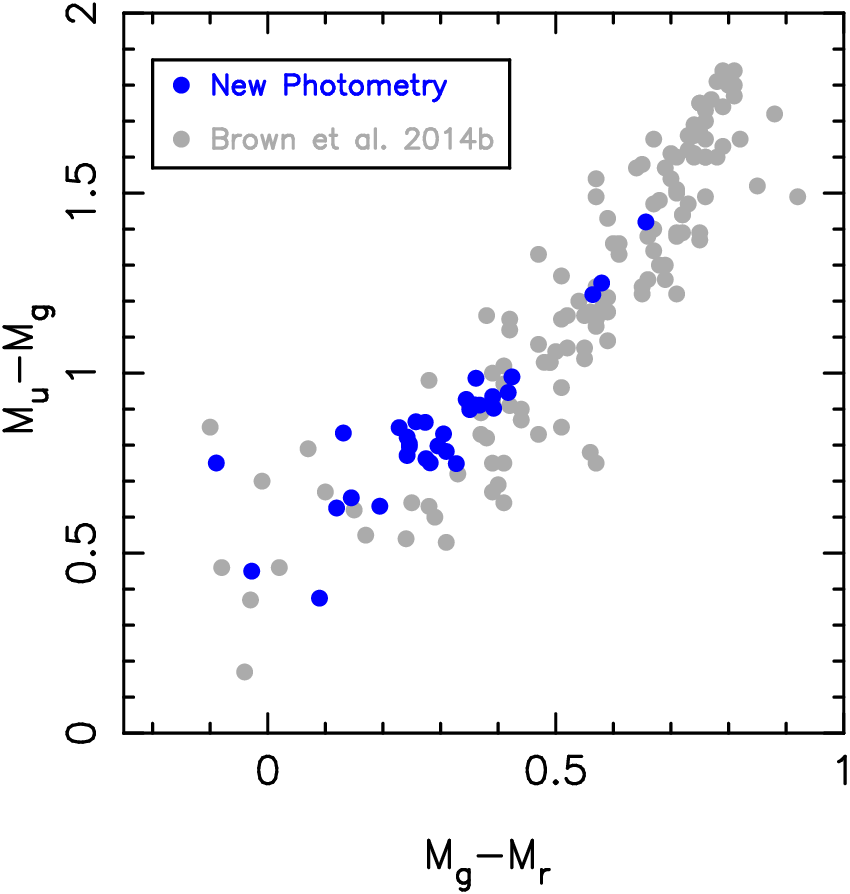}
\caption{Photometry of \citet{bro14} sample galaxies and galaxies from \citet{mou06} and \citet{mou10} with distances of less than $10~{\rm Mpc}$. As the photometric uncertainties are typically less than 0.1~mag., for the sake of clarity we have not included uncertainties in this plot (and this is the case for most plots in this paper). Unsurprisingly, the addition of nearby galaxies increases the number of blue low metallicity dwarfs in the sample.}
\label{fig:ugr}
\end{figure}

All galaxies in the \citet{bro14} sample with WISE colors of $W2-W3 \gtrsim 0$ (i.e., significant mid-infrared emission from warm dust), have low resolution 5--38$~{\rm \mu m}$ spectra from the {\it Spitzer} Infrared Spectrograph (IRS). The requirement for IRS spectra for star-forming galaxies was one of the biggest limitations on the \citet{bro14} sample size, and effectively excluded low luminosity dwarf galaxies from that sample. To correct for this weakness and extend our SFR calibration to low luminosities, we have added galaxies to the sample that have \citet{mou06} and \citet{mou10} drift-scan spectrophotometry, SDSS~III imaging and distances of less than $10~{\rm Mpc}$. Photometry for these galaxies was measured in the same bands as the \citet{bro14} sample (when available) and the optical color-color-diagram of the expanded sample of 161 galaxies is presented in Figure~\ref{fig:ugr}.

For each galaxy, the spectrophotometry was renormalized by a factor determined by dividing SDSS $g$-band aperture photon fluxes with $g$-band photon fluxes synthesized from the spectra. This resulted in systematic increases in the continuum and emission line fluxes of roughly 10\%, with larger corrections being common for galaxies brighter than $m_g=12$. Calibration of drift-scan spectrophotometry is non-trivial \citep[i.e.,][]{mou06,ken08} and for the brightest galaxies over-subtraction of the sky background may have enhanced the systematic errors. 

We expect some of the relationships presented in this paper to depend on total galaxy luminosity (or galaxy stellar mass), and these relationships can be non-linear. As a consequence, when calibrating SFR indicators we rescaled the broadband and emission line aperture fluxes by a factor equal to the $g$-band total flux divided by the $g$-band aperture flux. (This rescaling differs from a typical aperture bias correction, which accounts for broadband and emission line fluxes being measured using apertures of different sizes.) For most galaxies the total magnitude was the brighter of the aperture magnitude or the magnitude provided by the NASA-Sloan Atlas \citep{bla11}. For some galaxies where the aperture is smaller than the galaxy size and the NASA Sloan Atlas magnitude is absent or in error, we have remeasured ``total'' magnitudes using large aperture photometry\footnote{We remeasured total magnitudes for Mrk~33, NGC~337, NGC~628, NGC~2403, NGC~3049, NGC~3198, NGC~3351, NGC~3521, NGC~3627, NGC~4254, NGC~4559, NGC~4569, NGC~4656, NGC~4631, NGC~4670 and NGC~5055.}.

Radio continuum flux densities at $1.4~{\rm GHz}$ and $150~{\rm MHz}$ were determined using multiple sets of archival data. Our principal source of $1.4~{\rm GHz}$ flux densities is the NRAO VLA Sky Survey \citep[NVSS][]{con98}, which has an angular resolution of $45^{\prime\prime}$ and an RMS of $0.45~{\rm mJy}$ per beam. The NVSS flux calibration is tied to the \citet{baa77} absolute scale, and for compact sources NVSS flux densities agree with those of Westerbork/Einstein surveys to within a few percent \citep{con98}. Most of our galaxies have counterparts in default NVSS catalogue, but when available we used the flux densities from \citet{con02}, which includes single-dish flux densities for the brightest radio sources. A small number of galaxies have no catalogued NVSS flux densities and are relatively compact in size (less than $60^{\prime\prime}$ by $60^{\prime\prime}$), and for these galaxies we measured point source flux densities from the NVSS maps at the galaxy positions. 

Our principal source of $150~{\rm MHz}$ flux densities is the TIFR GMRT Sky Survey \citep[TGSS; e.g.,][]{bag11,gop12,sir14}, which has an angular resolution of $\sim 25^{\prime\prime}$ and an RMS of $\sim 3.5~{\rm mJy}$ per beam. We used the first alternative data release of the TGSS \citep[TGSS ADR1;][]{int16}, which provides images and catalogs for nearly the full TGSS survey area. TGSS ADR1 flux densities are tied to the \citet{sca12} scale, while comparisons with other surveys show TGSS flux densities for bright compact radio sources are 5\% brighter than 7C flux densities and almost identical to LOFAR flux densities \citep{int16}. 

To measure the TGSS flux densities for our galaxies, we defined elliptical apertures that encompassed the vast majority of the galaxy light identified in optical, mid-infrared and TGSS images. We then measured the flux densities directly from copies of the TGSS images with reduced angular resolution of  $\sim 45^{\prime\prime}$, which improves the detectability of extended emission. The TGSS ADR1 is optimized for imaging of compact sources, and therefore becomes less reliable for measuring flux densities for galaxies larger than a few arcminutes. For the brightest radio sources in our sample we used flux densities from the Sixth and Seventh Cambridge Surveys of Radio Sources \citep[6C, 7C;][]{6c1,6c2,6c3,6c4,6c5,6c6,7c} and GaLactic and Extragalactic All-Sky MWA Survey \citep[GLEAM;][]{way15,hur16}, which do not have the angular size limitations of the TGSS, but are more prone to source confusion. Changes to the selection criteria used for radio flux density measurements (e.g., the criteria used to exclude large galaxies) had little impact on our SFR indicator calibrations.

As relationships between SFR and luminosity can be non-linear, and many of our galaxies have distances of less than $10~{\rm Mpc}$, we utilize redshift independent distances (when available) or distances corrected for cosmic flows. Our sources of redshift independent distances are \citet{tul13} and \citet{sor14}, with the exception of NGC~4569 and UGCA~166, where we use distances from \citet{cor08} and \citet{mar10} respectively. For the nearest star forming galaxies, redshift independent distances are primarily from the tip of the red giant branch and cepheids, while beyond $10~{\rm Mpc}$ most redshift independent distances are derived from the Tully-Fisher relation. For the 72 galaxies without redshift independent distances, we use distances that account for cosmic flows induced by Virgo, the Shapley supercluster and the Great Attractor, using the prescription of \citet{mou00}.  Distance errors do not impact calibrations where SFR indicator luminosity is directly proportional to SFR. However, if the relationship between luminosity and SFR is a power-law with an index of 1.3, then a distance error of 20\% will translate to luminosity and SFR errors of 44\%, resulting in an offset from the power-law relation of $0.05~{\rm dex}$. This offset is relatively small, so we expect distance errors to have little impact on our SFR indicator calibrations. 


\section{Emission Line Fluxes}
\label{sec:emflux}

A significant change for this paper relative to previous studies using the \citet{mou06} and \citet{mou10} spectra is revised emission line fluxes. In order to minimize systematic differences in the emission-line fluxes from these two sources, we remeasured in a consistent way the strong nebular lines from the original flux-calibrated spectra. Following \citet{mou11}, we used modified versions of pPXF\footnote{\url{http://www-astro.physics.ox.ac.uk/~mxc/software/\#ppxf}} \citep{cap04} and {\sc gandalf}\footnote{\url{http://star-www.herts.ac.uk/~sarzi}} \citep{sar06} to model the stellar continuum and nebular emission lines respectively.  We fitted each stellar spectrum (after masking the emission lines) using a non-negative linear combination of ten Solar-metallicity \citet{bc03} population synthesis models with instantaneous-burst ages ranging from 5~Myr to 13~Gyr, assuming a \citet{cha03} IMF from $0.1-100~{\rm M_\odot}$.  

The fitting was executed twice, once using cross-correlation to allow for small adjustments to the fiducial redshift and a second time keeping the redshift fixed and fitting the continuum simultaneously with the stellar velocity dispersion.  We treated the selective extinction $E(B-V)$ as a free parameter for all the stellar ages and attenuate each spectrum using the \citet{cal00} dust law.  We verified that altering several of these assumptions had a negligible effect on our results: allowing a wider range of both sub- and super-Solar stellar metallicities; including a larger number of instantaneous-burst ages; adopting a different dust law \citep[e.g.,][]{odo94}; or allowing for time-dependent extinction \citep[e.g.,][]{cha00} changed the emission-line fluxes by ($<5\%$) in most cases.

Subtracting the best-fitting stellar continuum from the data resulted in a pure emission-line spectrum in which the Balmer and metal (forbidden) lines were optimally corrected for stellar absorption.  To measure the integrated emission-line fluxes, we simultaneously modeled the first four Balmer lines---H$\alpha$, H$\beta$, H$\gamma$, and H$\delta$---and the strong forbidden lines---[\ion{O}{2}]~$\lambda\lambda3726,3729$, [\ion{O}{3}]~$\lambda\lambda4959,5007$, [\ion{N}{2}]~$\lambda\lambda6548,6584$, and [\ion{S}{2}]~$\lambda\lambda6716,6731$--- assuming Gaussian line-profiles.  We carried this fitting out twice: on the first iteration we constrained the redshifts and intrinsic velocity widths of all the lines together and on the second iteration we relaxed these constraints and used the best-fitting parameters from the first iteration as initial guesses.  This second step was necessary because of uncertainties in the wavelength-dependent instrumental resolution and to account for any small ($<50$~km~s$^{-1}$) residual errors in the wavelength solution, particularly toward the edges of the spectra.

For galaxies with spectra from \citet{mou06} we find that our updated fluxes for the H$\alpha$ and H$\beta$ emission lines typically agree with the published fluxes to within $10\%$.  For galaxies with spectra from \citet{mou10}, the ${\rm H \alpha}$ emission-line fluxes are systematically lower by $\approx 20\%$ and the ${\rm H\beta}$ fluxes are higher by $\approx 10\%$ relative to the previously published values.  We attribute these non-negligible differences to an interpolation error in the spectra analyzed by \citet{mou10}. Finally, as noted in Section~\ref{sec:data}, spectrophotometry was renormalized by a factor determined by dividing SDSS $g$-band aperture photon fluxes with $g$-band photon fluxes synthesized from the spectra, which typically increased emission line fluxed by $\approx 10\%$. 

As the revisions to the ${\rm H \alpha}$ and ${\rm H \beta}$ emission lines fluxes were not negligible, we ran a series of cross checks to verify their accuracy. Visual inspection of plots was used to verify the accuracy of the stellar continuum subtract for each galaxy. Several diagnostic plots, including BPT diagrams and emission line ratios versus luminosity, had less scatter when revised emission line fluxes replaced published emission line fluxes. Finally, we cross checked the emission line fluxes against a simple model where the continuum was assumed to be constant near the relevant emission line, and found agreement to within $10\%$ for high equivalent width lines. Finally, the increase in emission line fluxes resulting from renormalizing the spectra with SDSS $g$-band photometry is consistent with offsets measured by \citet{ken08} when comparing \citet{mou06} spectra to narrow band imaging.

%

\section{SFR Indicator Calibration Sample}
\label{sec:sample}

Our SFR calibrations are anchored to Balmer decrement corrected ${\rm H\alpha}$ luminosities, so we excluded galaxies from the SFR calibration sample if the ${\rm H\alpha}$ or ${\rm H \beta}$ emission line fluxes had a signal-to-noise of less than five. The sample size does not strongly depend on the somewhat arbitrary choice of signal-to-noise ratio (many of the galaxies rejected by this threshold are passive ellipticals), but below this threshold ${\rm H\alpha}$ to ${\rm H\beta}$ flux ratios  often have uncertainties greater than one, resulting in highly uncertain Balmer decrement corrections. Our signal-to-noise threshold for ${\rm H\alpha}$ and ${\rm H \beta}$ reduced the sample from 161 galaxies to 109 galaxies, which are listed in Table~\ref{table:eflux}.

\begin{figure}
\plotone{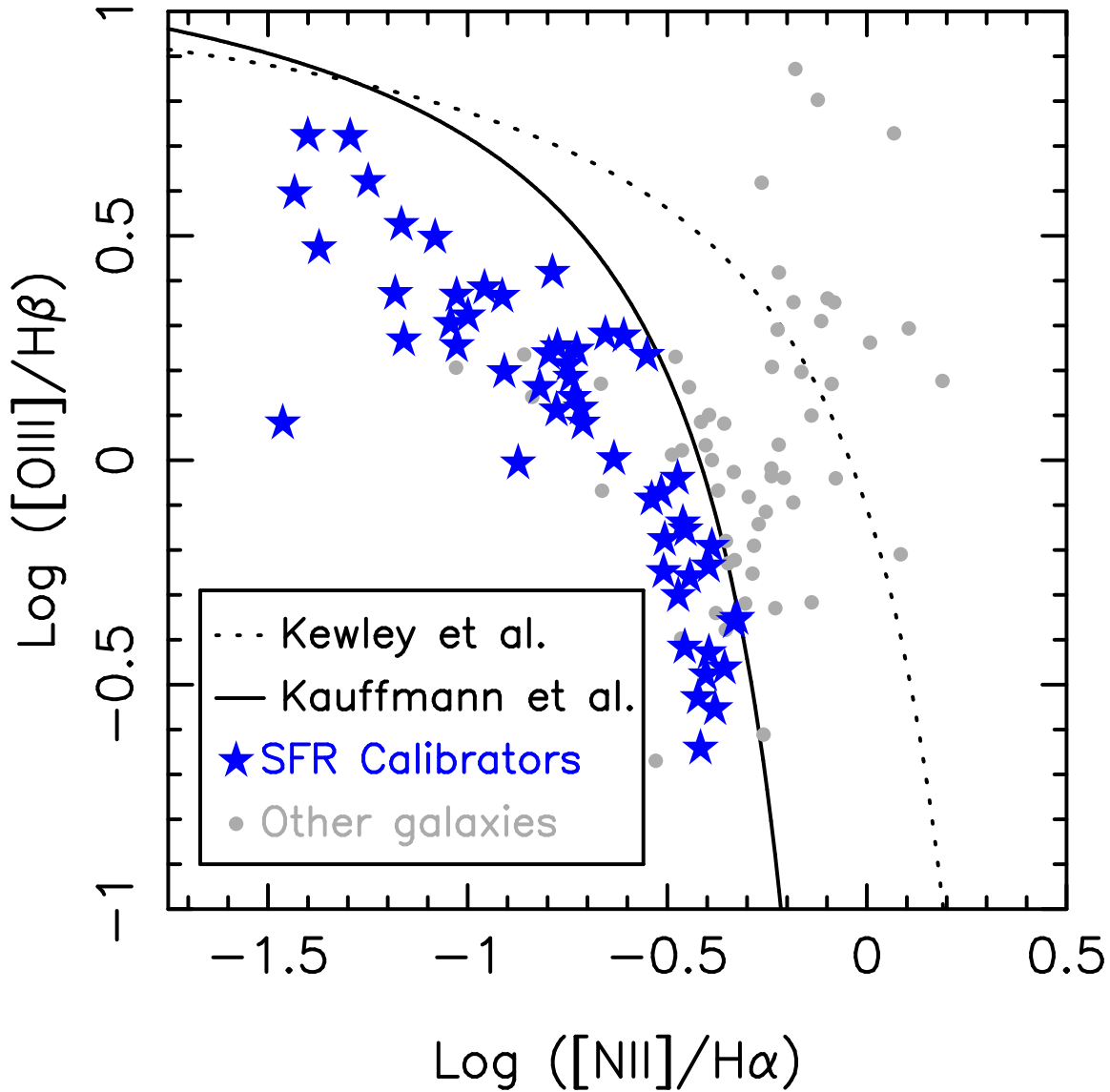}
\caption{The BPT diagram for galaxies in the \citet{bro14} sample. The spectral classification criteria of \citet{kew01} and \citet{kau03} are also plotted, and these were used to classify galaxies as star forming galaxies, AGNs and potential composite objects. Blue stars show galaxies in the SFR calibration sample while grey dots denote other galaxies, including those with low signal-to-noise emission line measurements.}
\label{fig:bpt}
\end{figure}

The \citet{bro14} sample includes LINERS and AGNs where ${\rm H\alpha}$ emission is not the result of star formation. As we illustrate in Figure~\ref{fig:bpt}, we excluded these galaxies from the SFR indicator calibration sample using the BPT diagram \citep{bal81} and the criterion of \citet{kau03}. We also considered excluding AGNs identified using the mid-infrared color criterion of \citet{ste05}, but this criterion also excludes some low metallicity dwarf galaxies that we wish to keep in the sample. Finally, as we wanted our SEDs to be representative of entire galaxies, we excluded galaxies from the SFR calibration sample if the $g$-band aperture and total magnitudes differed by more than 0.75~mag. Thus, by construction, we expect our relations derived from entire galaxies will differ from those using subregions of galaxies and H~II regions \citep[e.g.,][]{cal07,rel07,ken09}. Our criteria reduced our final SFR indicator calibration sample to 66 galaxies, although for any given calibration less galaxies are used due to data coverage and signal-to-noise limitations.

\begin{figure}   
\plotone{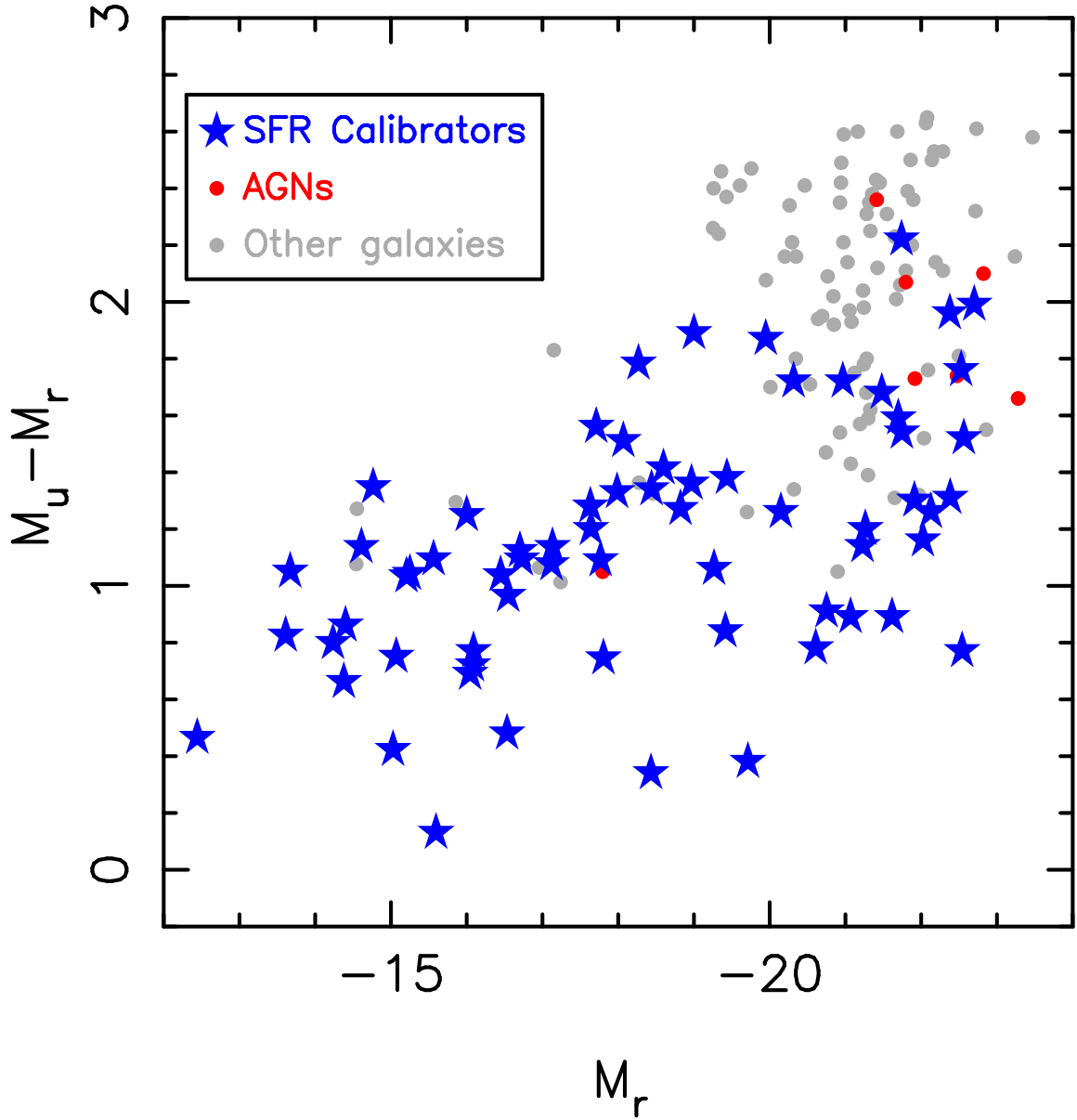}
\caption{SDSS optical color-magnitude diagram for the sample. Galaxies in the SFR indicator calibration sample are shown with blue stars, BPT selected AGNs are denoted by red circles and other galaxies are shown in grey (including galaxies with low signal-to-noise emission line fluxes). The SFR indicator calibration galaxies span a broad range of optical color and absolute magnitude.}
\label{fig:optical}
\end{figure}

The optical color-magnitude diagram of the \citet{bro14} sample and the SFR indicator calibration sample are provided in Figure~\ref{fig:optical}. The SFR indicator calibration sample spans $-24<M_r<-12$ and $-0.3<u-r<2.3$, and includes galaxies with optical colors approaching those of passive galaxies. This broad distribution of optical properties reflects the deliberate targeting of galaxies spanning a broad range of optical properties by \citet{mou06} and \citet{mou10}.

\begin{figure*}
\plottwo{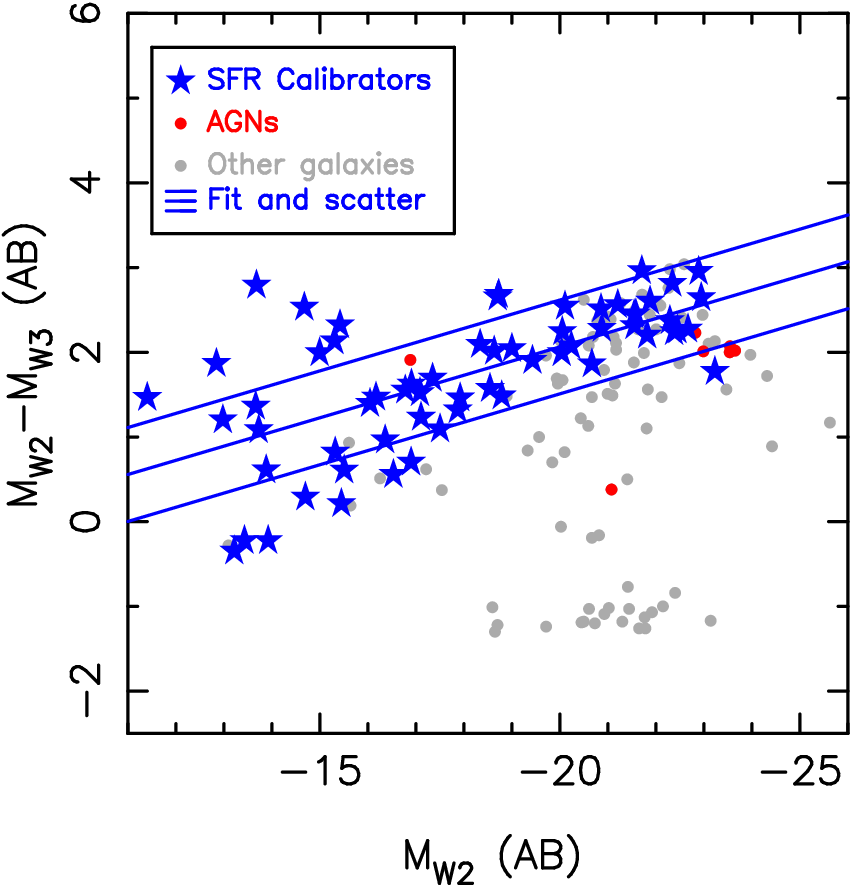}{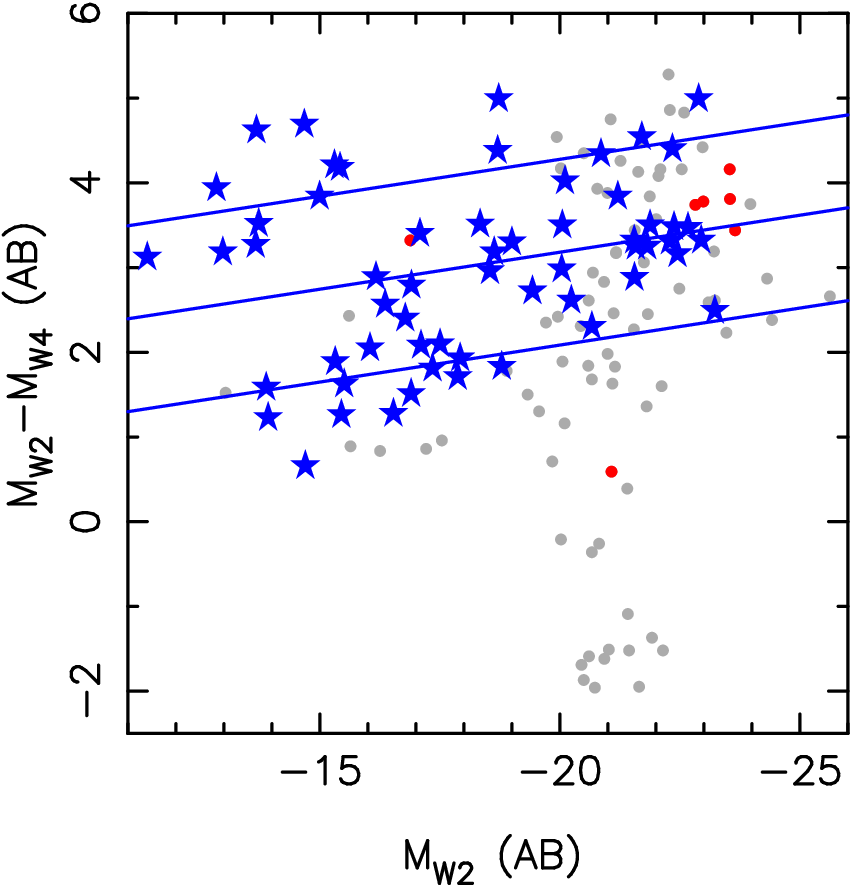}
\caption{WISE mid-infrared color-magnitude diagrams for the sample. Compared to the optical color-magnitude diagram, SFR indicator calibration galaxies are clearly separated from the locus of passive galaxies (located at the bottom right of both panels). While both WISE $W3$ and $W4$ luminosities are used as SFR indicators, the widths of the $M_{W2}-M_{W3}$ and $M_{W2}-M_{W4}$ distributions differ considerably from each other, and this may imply $W3$ has a limited dynamic range as a SFR indicator.}
\label{fig:mid}
\end{figure*}

We plot the mid-infrared color-magnitude diagrams of the sample in Figure~\ref{fig:mid}, and this figure provides several reasons for caution when using SFR indicators. Unlike the optical color-magnitude diagram, there is a significant gap between the SFR indicator calibration sample and passive galaxies. Several of the galaxies that fall between the star forming and passive loci are forming stars, but their spectra do not meet the criteria for inclusion in the SFR calibration sample. For example, NGC~3190 and NGC~4725 both lack detectable H${\rm \beta}$ emission in their drift scan spectra, but both show clear evidence for star formation in GALEX images and SINGS continuum subtracted H${\rm \alpha}$ images \citep{ken03}. Our SFR indicator calibration sample does not probe the lowest specific star formation rates (sSFRs), and this may be true of other calibrations in the literature that have similar limitations. 

At fixed stellar mass, one may expect different SFR indicators to have comparable logarithmic luminosity ranges, but this is not the case for the WISE $W3$ and $W4$ bands. Figure~\ref{fig:mid} illustrates that the distributions of $W3$ and $W4$ luminosities at fixed $W2$ absolute magnitude (or approximate stellar mass) differ considerably from each other. When we fit to the mid-infrared color-magnitude relations for the SFR calibration sample, we find both relations are tilted and the data show significant scatter about these relations, which is to be expected as mid-infrared luminosity is not a linear function of SFR \citep[e.g.,][]{lee13,cat15}, sSFR won't necessarily be constant with stellar mass and the star-forming ``main sequence'' has significant scatter at fixed mass.  The $1\sigma$ scatter for $M_{W2}-M_{W3}$ colors about the best fit relation is $\sim 0.6~{\rm mag}$, which is considerably less than the $1\sigma$ scatter for $M_{W2}-M_{W4}$ colors data, which is $\sim 1~{\rm mag}$. As the sSFRs derived from $H{\rm \alpha}$ luminosities span approximately an order of magnitude, the relatively narrow range of $M_{W2}-M_{W3}$ colors may imply that WISE $W3$ has a limited dynamic range as a SFR indicator. Furthermore, galaxies in the SFR calibration sample have colors that span $0.0<M_{W3}-M_{W4}<2.3$, so in many instances SFRs determined with the WISE $W3$ and $W4$ bands will differ significantly from each other. 

\section{Star Formation Rate Indicator Calibrations}
\label{sec:calibration}

Our SFR indicator calibrations are anchored to Balmer decrement corrected ${\rm H\alpha}$ luminosities assuming a \citet{fit99} dust attenuation curve with $R_V=3.1$ and Case B recombination with an effective temperature of $10,000~{\rm K}$ and $n_e=10^2~{\rm cm}^{-3}$, where the ratio of ${\rm H\alpha}$ luminosity to ${\rm H\beta}$ luminosity is 2.86 \citep{sto95,dop03}. This choice is transparent and easier to replicate than more complex modeling of galaxy SEDs and and dust geometry, but its simplifying assumptions must be wrong in detail (e.g., obscuration by a dusty screen). 

The assumptions we used when determining Balmer decrement corrected ${\rm H\alpha}$ luminosities probably have limited impact on SFR calibrations, and this is discussed in detail by \citet{ken09}.  For example, \citet{cal07} found that attenuations for ${\rm H\alpha}$ determined using the Balmer decrement technique show no systematic offset relative to those determined with ${\rm Pa \alpha / H \alpha}$ ratios. Furthermore, when we fitted models to relationship between SFR indicator luminosity and Balmer decrement corrected ${\rm H\alpha}$ luminosity, we found the parameter values changed by $\lesssim 2\sigma$ when we substituted a \citet{cal00} dust attenuation law for our default \citet{fit99} dust attenuation law.

\begin{figure*}
\plottwo{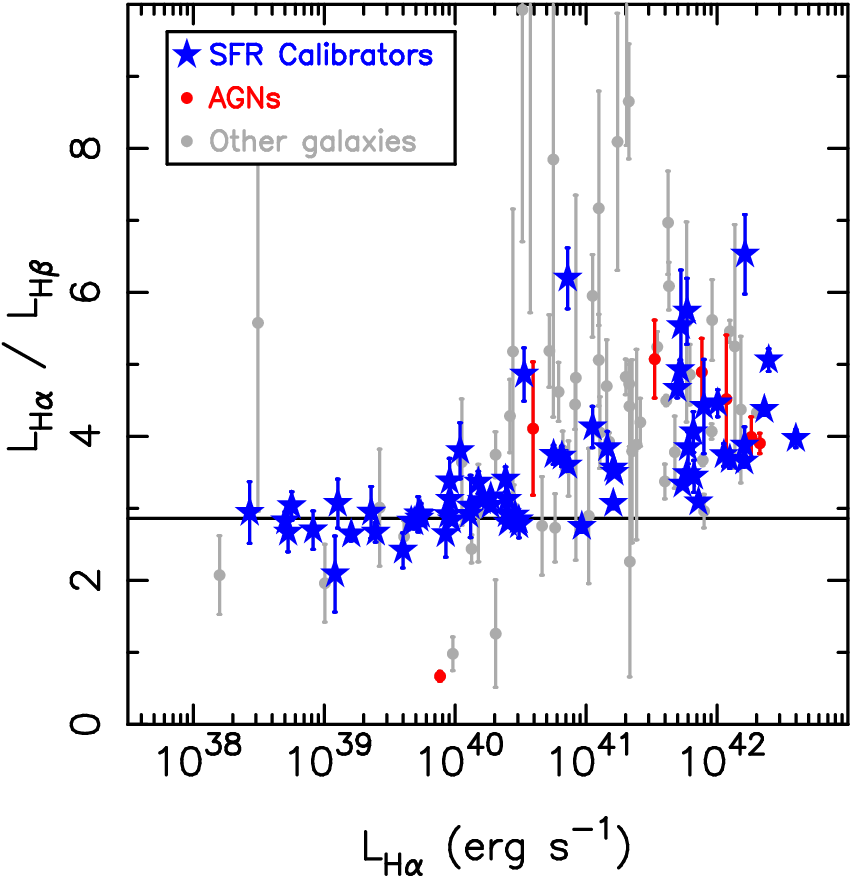}{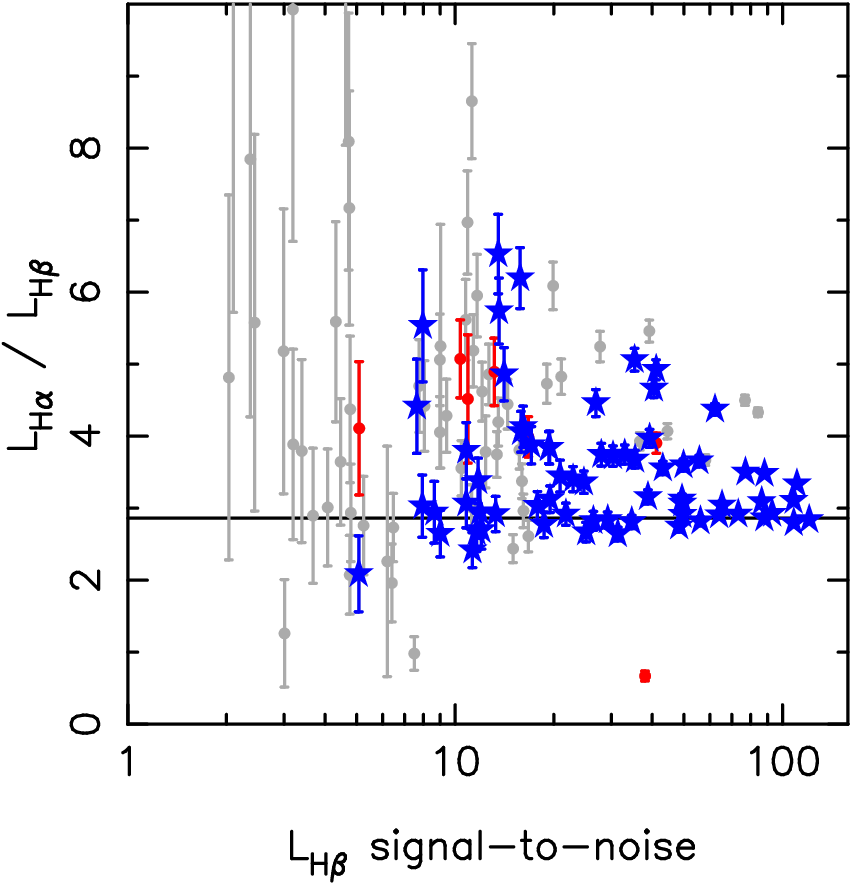}
\caption{The ratio of observed H$\alpha$ luminosity to observed H$\beta$ luminosity, as a function of H$\alpha$ luminosity (left panel) and H$\beta$ signal-to-noise (right panel). Galaxies used for the star formation rate calibration are shown with blue stars, BPT selected AGNs are shown with red dots and other galaxies (including those with low signal-to-noise emission line fluxes) are shown in grey. Dust obscuration increases with increasing luminosity, while at low luminosities the ratio of $\alpha$ luminosity to H$\beta$ luminosity approaches the value expected for Case~B recombination. The spread of H$\alpha$ luminosity to H$\beta$ luminosity ratios does depend on signal-to-noise, with spuriously low values being associated with mediocre signal-to-noise.}
\label{fig:ratio}
\end{figure*}

In Figure~\ref{fig:ratio} we plot the ratio of the H$\alpha$ to H$\beta$ flux as a function of H$\alpha$ luminosity, along with the expected ratio for $10,000~{\rm K}$ Case B recombination. The value of ${\rm H\alpha}$ luminosity divided by ${\rm H\beta}$ luminosity for Case B recombination can vary from 2.75 to 3.04 for temperatures ranging from $20,000~{\rm K}$ to $5,000~{\rm K}$, but we do not expect this source of error to dominate the observed scatter in SFR indicator calibrations. As has been reported in the prior literature \citep[e.g.,][]{lee09}, blue compact dwarf galaxies that have low ${\rm H\alpha}$ luminosities (but high sSFRs) also have relatively little dust obscuration, and the H$\alpha$ to H$\beta$ flux ratios asymptote towards the expected range for Case B recombination. 

\begin{figure}
\plotone{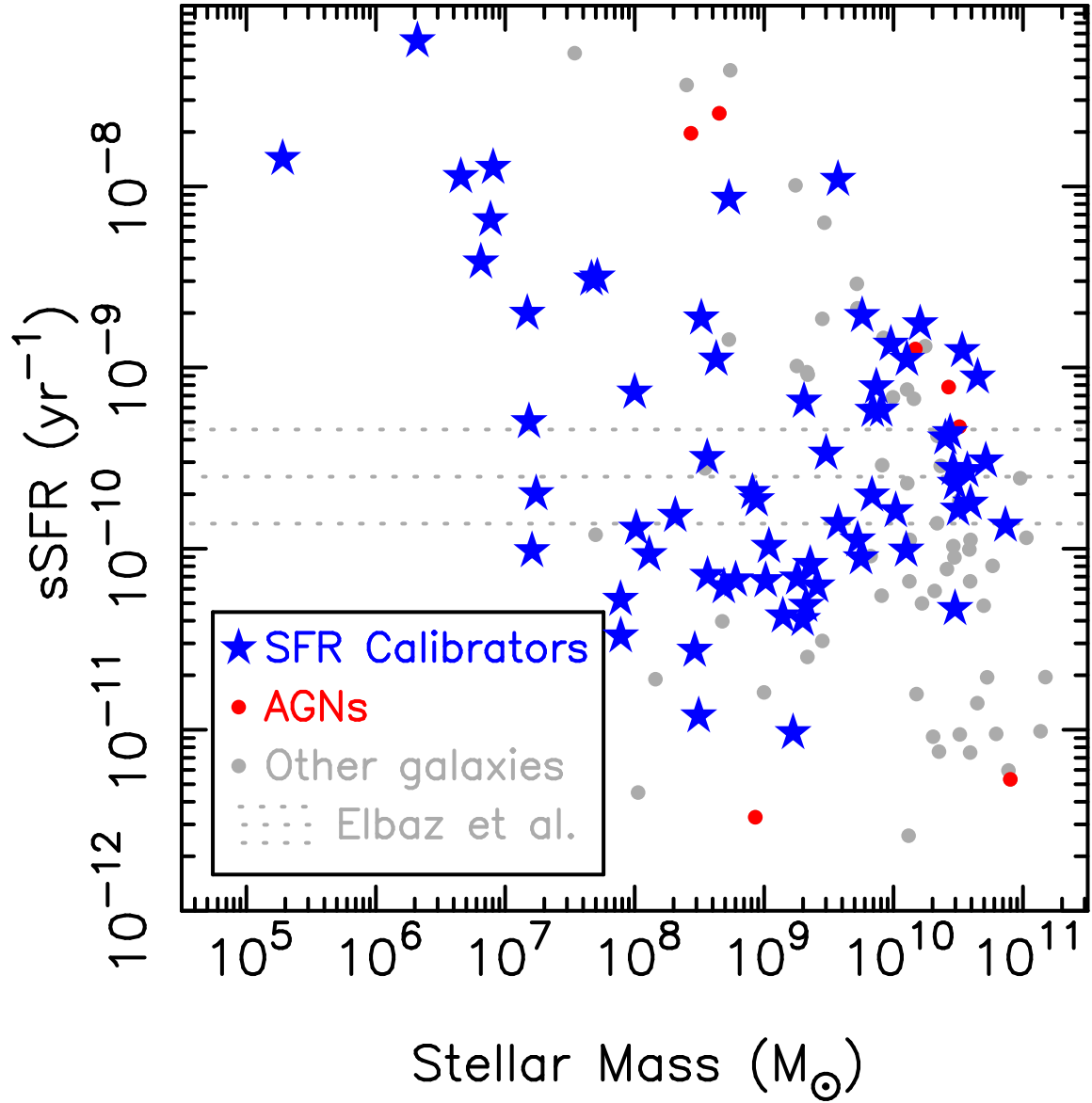}
\caption{sSFR as a function of galaxy mass. sSFRs decrease with increasing stellar mass, and at fixed stellar mass the sSFRs have a range of two orders of magnitude. The location of the ``star forming main sequence'' is illustrated with the 16th, 50th and 84th percentiles from \citet{elb11}. The ``star forming main sequence'' is not particularly evident in our sample, which is an artifact of the sample selection, which had an emphasis on spanning parameter space \citep{mou06,mou10}.}
\label{fig:ssfr}
\end{figure}

Figure~\ref{fig:ssfr} shows the sSFRs of the sample galaxies as a function of their stellar mass. SFRs were determined using  
\begin{equation}
SFR({\rm M_\odot~ yr^{-1}}) =  5.5\times 10^{-42} L_{\rm H\alpha} ({\rm erg~s^{-1}})
\end{equation}
\citep{ken09}, which uses a \citet{kro01} IMF and a constant SFR. Approximate stellar masses were determined using WISE $W1$ and $W2$ photometry and the relation of \citet{clu14}, with the addition of 0.07~dex to convert from a \citet{cha03} IMF to a \citet{kro01} IMF. sSFRs decrease with increasing stellar mass, and at fixed stellar mass the sSFRs have a range of two orders of magnitude. The ``star forming main sequence'' \citep[e.g.,][]{noe07,elb11} is not particularly evident in Figure~\ref{fig:ssfr}, which is an artifact of the sample selection, which emphasized spanning parameter space rather than providing a flux limited galaxy sample \citep{mou06,mou10}.

For consistency with (much of) the prior literature, we use powers in units of ${\rm W~Hz^{-1}}$ for the radio continuum and $\nu L_\nu$ in units of ${\rm ergs~s^{-1}}$ for the ultraviolet and mid-infrared, where the frequency $\nu$ is determined from the effective wavelength of the relevant filter. In the ultraviolet and mid-infrared the flux density is given by 
\begin{equation}
f_\nu = 3631 {\rm Jy} \times 10^{-0.4 m}.
\end{equation}
where $m$ is the AB apparent magnitude. We caution that some flux densities presented in the literature do not use this definition, and this can result in systematic offsets of several percent. The effective wavelengths of the relevant filters are presented in Table~\ref{table:filters}. The effective wavelength depends on the weighting function used, corresponding to the assumed spectrum of the source being observed, so we choose to use effective wavelengths as published by the relevant survey/satellite teams. For the calibration of radio continuum as a SFR indicator we used the flux densities from NVSS and TGSS ADR1 \citep{con98,con02,int16}, and frequencies of $1.40~{\rm GHz}$ or $150~{\rm MHz}$. 

To model the relationship between SFR indicator luminosity and Balmer decrement corrected H$\alpha$ luminosity, we have used two parameterizations. The first is a power-law where the index and normalization are free parameters, which is commonly used and thus simplifies direct comparisons with the prior literature. Table~\ref{table:litcal} provides an incomplete list of power-law SFR calibrations from the prior literature, including models with power-law indices fixed at one \citep[e.g.,][]{ken09}. Table~\ref{table:litcal} provides at least four calibrations for each filter, with an emphasis on calibrations based on ${\rm H\alpha}$ and ${\rm Pa \alpha}$, which aids direct comparison with our work\footnote{Please note Table~\ref{table:litcal} does not include some calibrations that utilize total infrared luminosity \citep[e.g.,][]{got11,ruj13} and some papers listed in Table~\ref{table:litcal} use several different calibration methods \citep[e.g.,][]{rie09,dav16}}. To simplify comparisons of different models, we have rewritten the parameterizations from the prior literature so they are a function of ${\rm H\alpha}$ luminosity with the normalization being the SFR indicator luminosity of a galaxy with an ${\rm H\alpha}$ luminosity of $10^{40}~{\rm erg~s^{-1}}$.

The power-law parameterization assumes two galaxies with the same SFR but very different masses and metallicities will have the same SFR indicator luminosity, which may not necessarily be the case. For example, we may expect a metal rich $L^*$ galaxy will have higher dust content and higher mid-infrared luminosity at a given SFR than a metal poor dwarf galaxy with the same SFR. This motivated our second parameterization of the relationship between SFR indicator luminosity and SFR. 

Our second parameterization assumes that SFR indicator luminosity is directly proportional to SFR for galaxies of a given mass, with the normalization being a power-law function of galaxy mass. To simplify the use of this parameterization, we have used {\it Spitzer} $4.5~{\rm \mu m}$ and WISE $W2$ luminosities as stellar mass proxies \footnote{Although the {\it Spitzer} $4.5~{\rm \mu m}$ and WISE $W2$ bands include ${\rm Br\alpha}$, for most star forming galaxies the ${\rm Br \alpha}$ emission line fluxes \citep[e.g.,][]{ima10} are small compared to the {\it Spitzer} and WISE broadband fluxes.}. This parameterization has the same number of free parameters as power-law models, but may be less prone to error when extrapolated to high and low SFRs if its underlying assumption is valid (i.e., luminosity is a linear function of SFR for galaxies of a given mass). 


For each relation, the $1\sigma$ scatter of the data about the best-fit was determined by finding the scatter that encompassed 68\% of the data, and any galaxies more than $2\sigma$ from the best-fit relation were flagged as potential outliers. Wide-field surveys cannot always apply stringent BPT criteria, so we also present measurements of the scatter using galaxies that meet the less stringent BPT criterion of \citet{kew01}. This second measurement of the scatter may overestimate the scatter for magnitude limited samples, as AGNs and LIRGs are over-represented in the \citet{bro14} sample. Parameter values are presented for galaxies with H${\rm \alpha}$ luminosities of $10^{40}~{\rm erg~s^{-1}}$ (rather than extrapolating to $1~{\rm erg~s^{-1}}$) to reduce quoted uncertainties.

\begin{figure}
\plotone{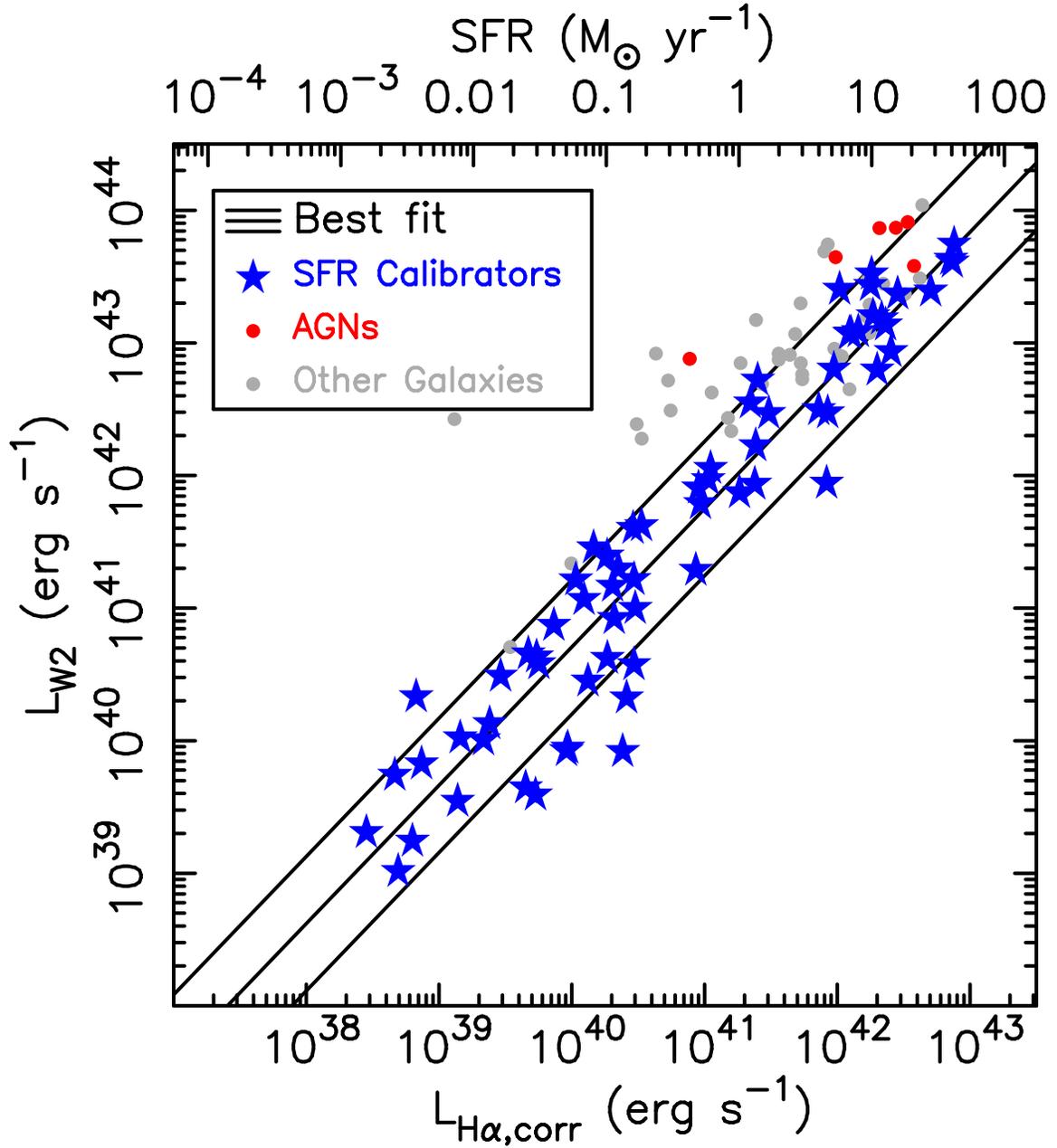}
\caption{WISE $W2$ ($4.6~{\rm \mu m}$) luminosity as a function of Balmer decrement corrected H$\alpha$ luminosity. A power-law fit to the data, and the $\pm 1 \sigma$ scatter of the data, is shown with black lines. As $W2$ is a better tracer of stellar mass than SFR, this plot illustrates luminosity-luminosity correlations in the sample. Unlike fits to data at longer wavelengths, the best fit power-law has an index close to one while the scatter of the data around the fit is relatively large (0.4~dex).}
\label{fig:w12}
\end{figure}

As a sobriety test for the relations presented in this paper, in Figure~\ref{fig:w12} we present W2 ($4.6~{\rm \mu m}$) luminosity as a function of Balmer decrement corrected H$\alpha$ luminosity. Although WISE $W2$ is usually a proxy for stellar mass rather than SFR, near-infrared luminosity does depend on stellar population age \citep[e.g.,][]{bc03} and it thus isn't entirely independent of SFR. The power-law fit to the WISE $W2$ data has an index close to one and the scatter around the best-fit power-law is 0.4~dex, which is smaller than the scatter seen in sSFR versus stellar mass for our sample (illustrated by Figure~\ref{fig:ssfr}). Galaxies with lower sSFRs than the BPT selected calibration sample fall to the left of the power-law fit, having significant WISE $W2$ emission but low SFRs. We remind adventurous readers to not use WISE $W2$ as an SFR indicator. 

\subsection{Ultraviolet}

To use $FUV$ as a SFR indicator, one must model the dust extinction and the intrinsic SED of the galaxy stellar population. While one can model entire SEDs to derive stellar populations and dust extinction \citep[e.g.,][]{dac08,nol09} this isn't always practical for wide-field surveys (e.g., much of the southern sky currently lacks $ugriz$ imaging while 2MASS $JHK_S$ imaging is shallow). As $NUV$ imaging is almost always available with $FUV$ imaging, we have adopted corrections for dust extinction corrections that are a function of $M_{FUV}-M_{NUV}$ color. This effectively makes our $FUV$ calibrations composites with $NUV$, whereas monochromatic calibrations are available for all the other bands presented in this paper.

\begin{figure*}
\plottwo{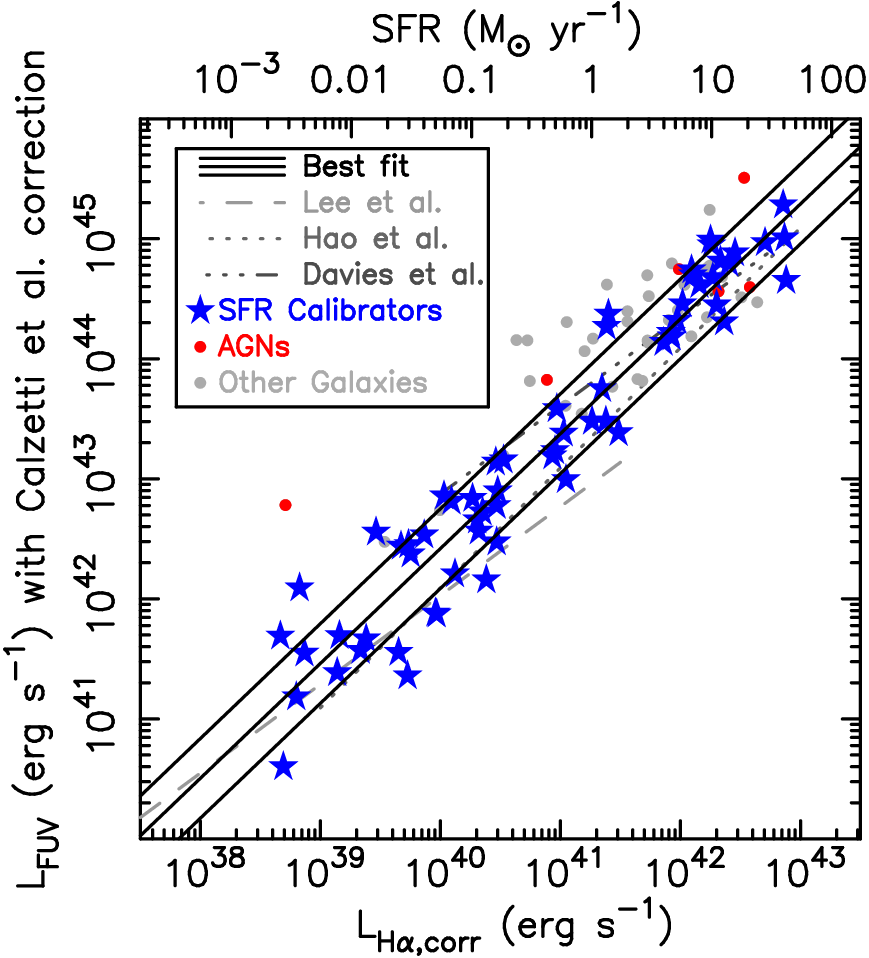}{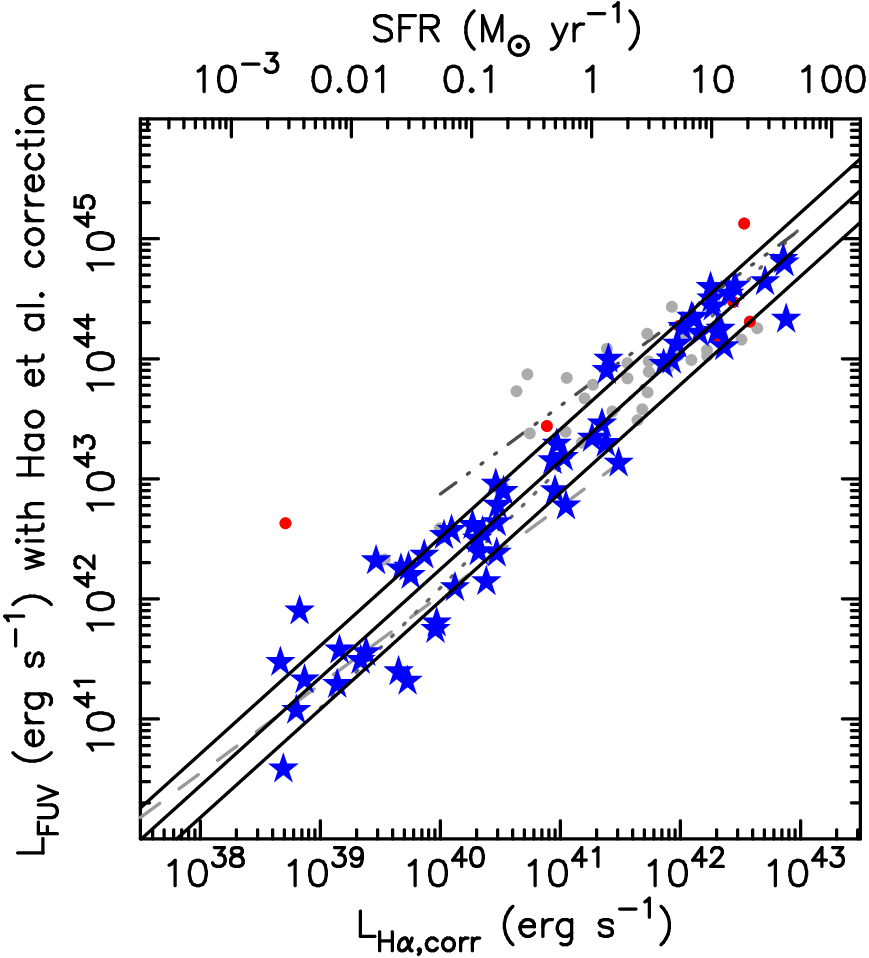}
\caption{Dust obscuration corrected GALEX $FUV$ luminosity as a function of Balmer decrement corrected $H\alpha$ luminosity, with a \citet{cal00} and \citet{hao11} corrections for dust obscuration (derived from observed $M_{FUV}-M_{NUV}$) used in the left and right panels respectively.  A STARBURST99 \citep{lei99} model for a 100~Myr old stellar population with Kroupa IMF \citep{hao11} is comparable to the fits to our data. While the power-law indices are within 10\% of the expected value of one, the scatter of the data around the fits is $\sim 0.3~{\rm dex}$ for both panels.}
\label{fig:fcal}
\end{figure*}

In Figure~\ref{fig:fcal} we present two $FUV$ calibrations that use different stellar population and dust extinction corrections. In the left panel of Figure~\ref{fig:fcal} we have assumed the stellar population spectrum of star-forming galaxies has a dust free color of $M_{FUV}-M_{NUV}=0$, which is comparable to the bluest galaxies in our sample and young populations \citep[e.g.,][]{gil07,lis08}, and then corrected for internal dust extinction using a \citet{cal00} extinction law. In the right panel of Figure~\ref{fig:fcal} we have assumed the stellar population spectrum of star-forming galaxies has a dust free color of $M_{FUV}-M_{NUV}=0.022$ \citep{hao11} and we have used the empirical model of $FUV$ dust attenuation as a function of $M_{FUV}-M_{NUV}$ from \citet{hao11}. Both dust corrections make assumptions about stellar populations and dust obscuration that must be wrong for many individual star forming galaxies, but (as we discuss below) the impact of these assumptions is reduced via empirical calibration of $FUV$ with ${\rm H\alpha}$.

In Figure~\ref{fig:fcal} we present dust corrected GALEX $FUV$ luminosity as a function of Balmer decrement corrected ${\rm H\alpha}$ luminosity. Power-law fits to the data are also plotted in Figure~\ref{fig:fcal}, and the relevant parameter values provided in Table~\ref{table:calibration}. Both fits have power-law indices within 10\% of the expected value of one, and the fits are comparable to the predicted relationship between $FUV$ and ${\rm H\alpha}$ from STARBURST99 \citep{lei99} for a 100~Myr old stellar population with a Kroupa IMF \citep{hao11}. As the power-law fits have indices close to one, we have not attempted to use our alternative parameterization to calibrate the $FUV$ data. Empirical relations for GALEX $FUV$ luminosity as a function of ${\rm H\alpha}$ luminosity \citep{lee09,dav16,jai16} show significant offsets with respect to each other and our work, and this may be partially explained by different models for correcting dust attenuation. Unfortunately the scatter of the data around our best fit power-laws is $\sim 0.3~{\rm dex}$, and thus not much better than what was achieved with WISE $W2$. 

\subsection{Mid-infrared}

Mid-infrared emission from star forming galaxies is dominated by the blackbody radiation from warm dust and emission features attributed to PAHs, and thus mid-infrared emission resulting from star formation has dependencies on dust content (and thus metallicity), geometry and temperature. Furthermore, the mid-infrared emission from galaxies can include contributions from dust heated by old stellar populations (``galactic cirrus''), AGNs and the Rayleigh-Jeans tail of stellar spectra. Mid-infrared emission from galaxies is thus the result of complex astrophysics, and it is a fortunate accident that the relationship between star formation and mid-infrared luminosity can be empirically modeled with relatively simple functions \citep[e.g.,][]{wu05,ken09,cat15}.

\begin{figure*}
\plottwo{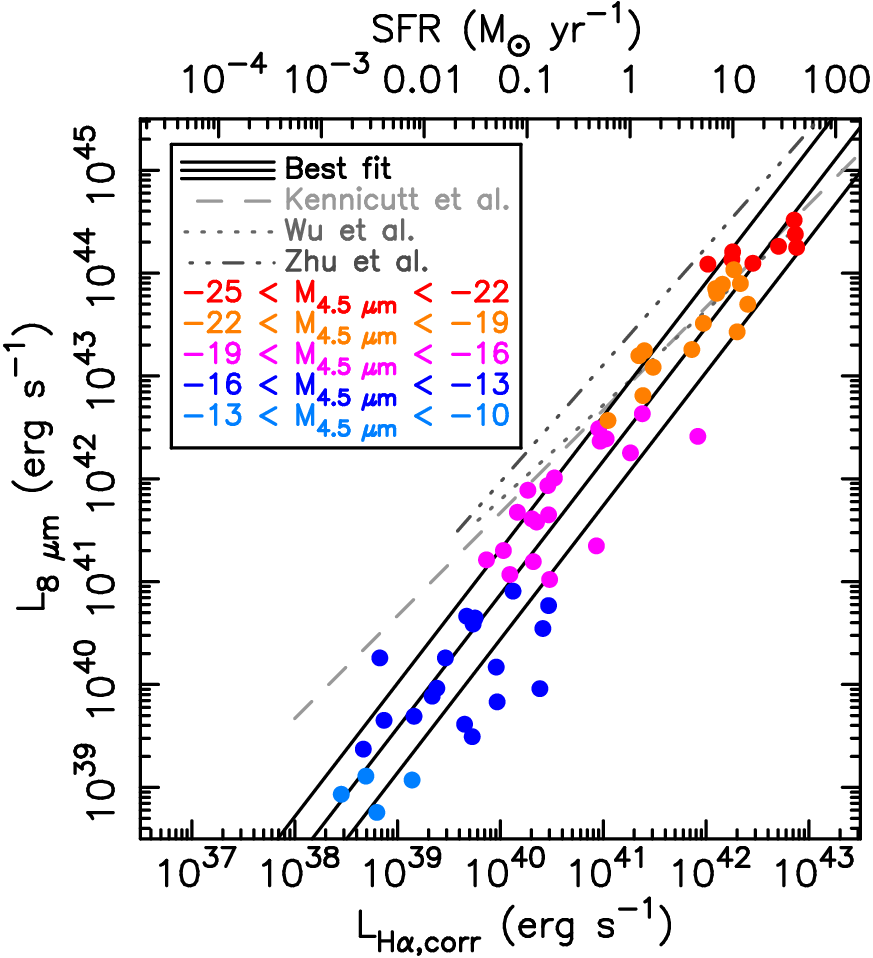}{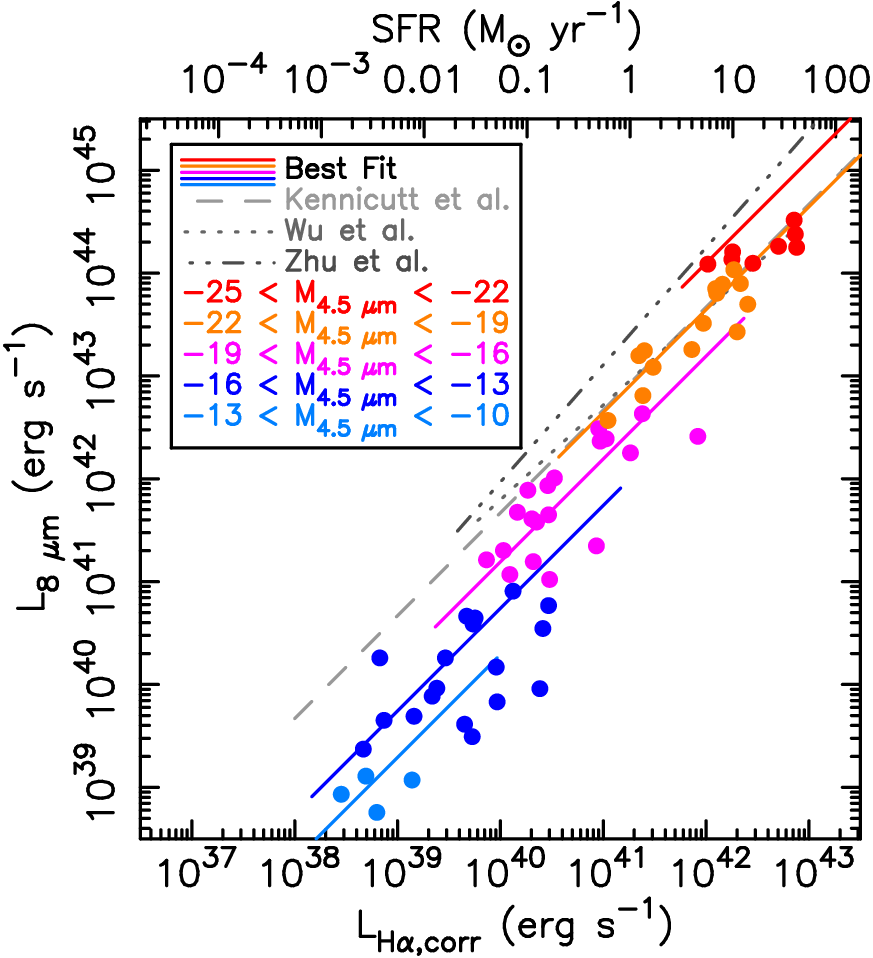}
\caption{{\it Spitzer} $8~{\rm \mu m}$ luminosity as a function of Balmer decrement corrected H$\alpha$, with data points color coded by $4.5~{\rm \mu m}$ absolute magnitude (a rough stellar mass proxy). In the left panel we plot a power-law fit to the data, while in the right panel we plot a fit where $8~{\rm \mu m}$ luminosity scales linearly with SFR and normalization is a function of $4.5~{\rm \mu m}$ luminosity. While our power-law fit has an index of $1.30\pm 0.05$, power-laws from the prior literature have indices closer to one.}
\label{fig:I4cal}
\end{figure*}

\begin{figure*}
\plottwo{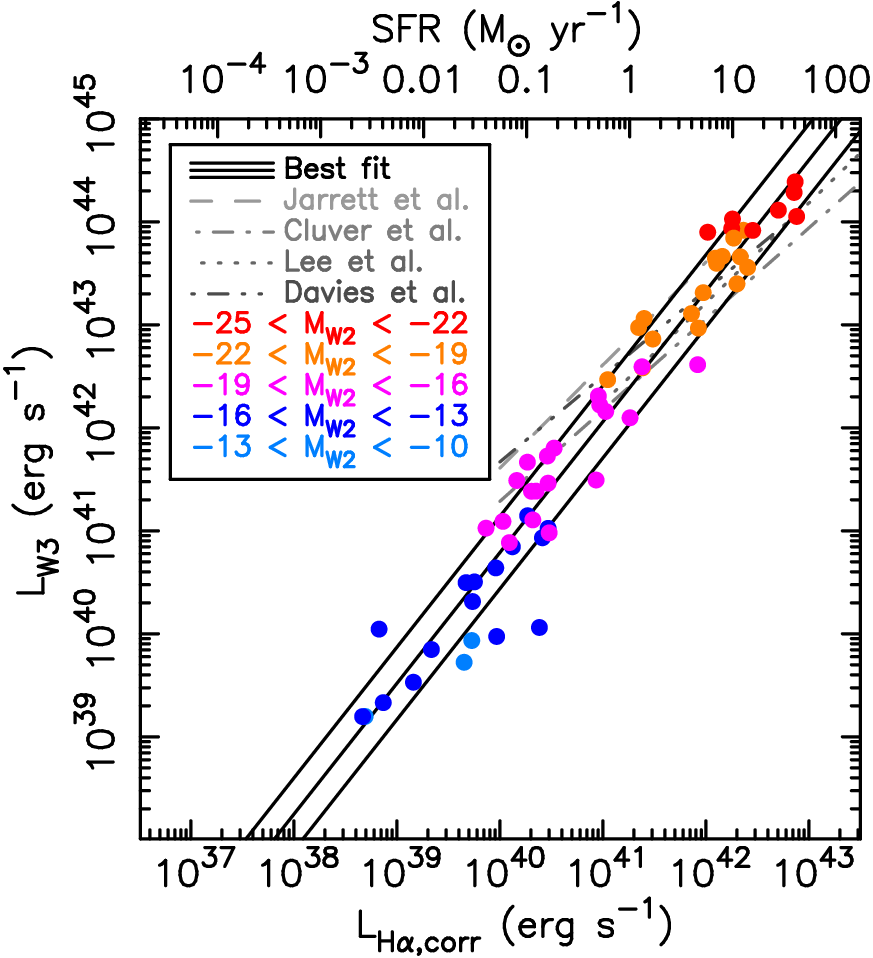}{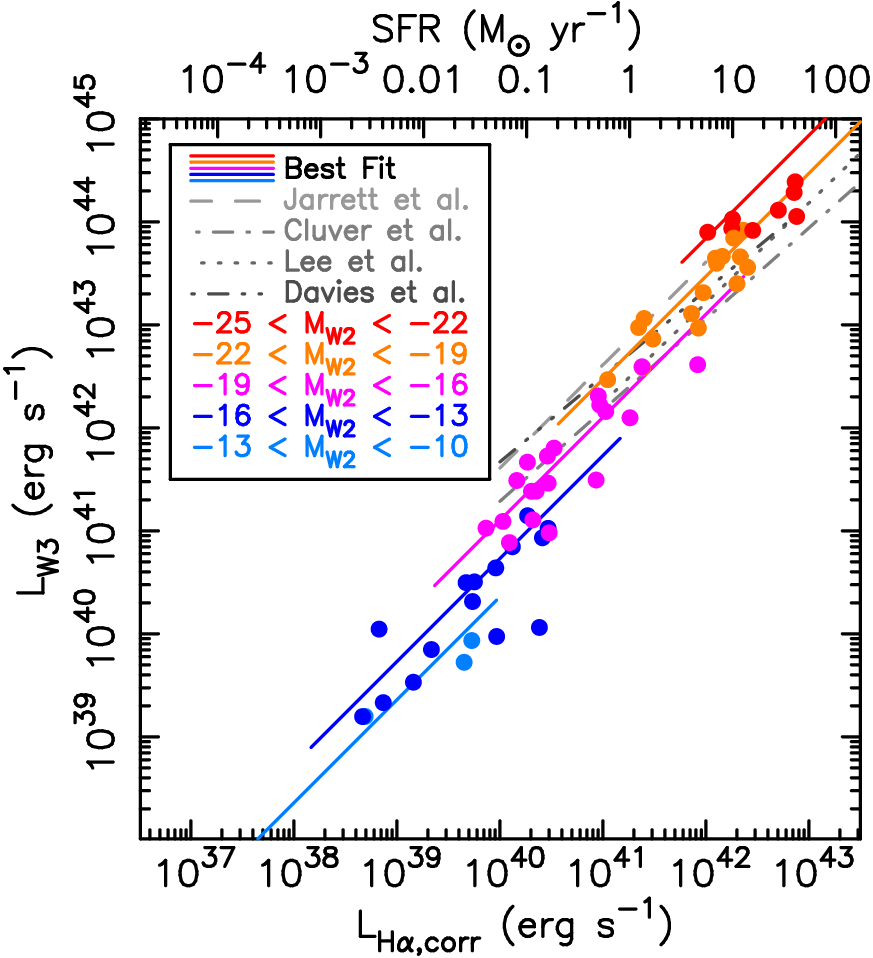}
\caption{WISE $W3$ luminosity as a function of Balmer decrement corrected H$\alpha$.  For dwarf galaxies, we measure systematically higher H$\alpha$ luminosities and SFRs at fixed $W3$ luminosity relative to extrapolations of relations from the prior literature.}
\label{fig:w3cal}
\end{figure*}

\begin{figure*}
\plottwo{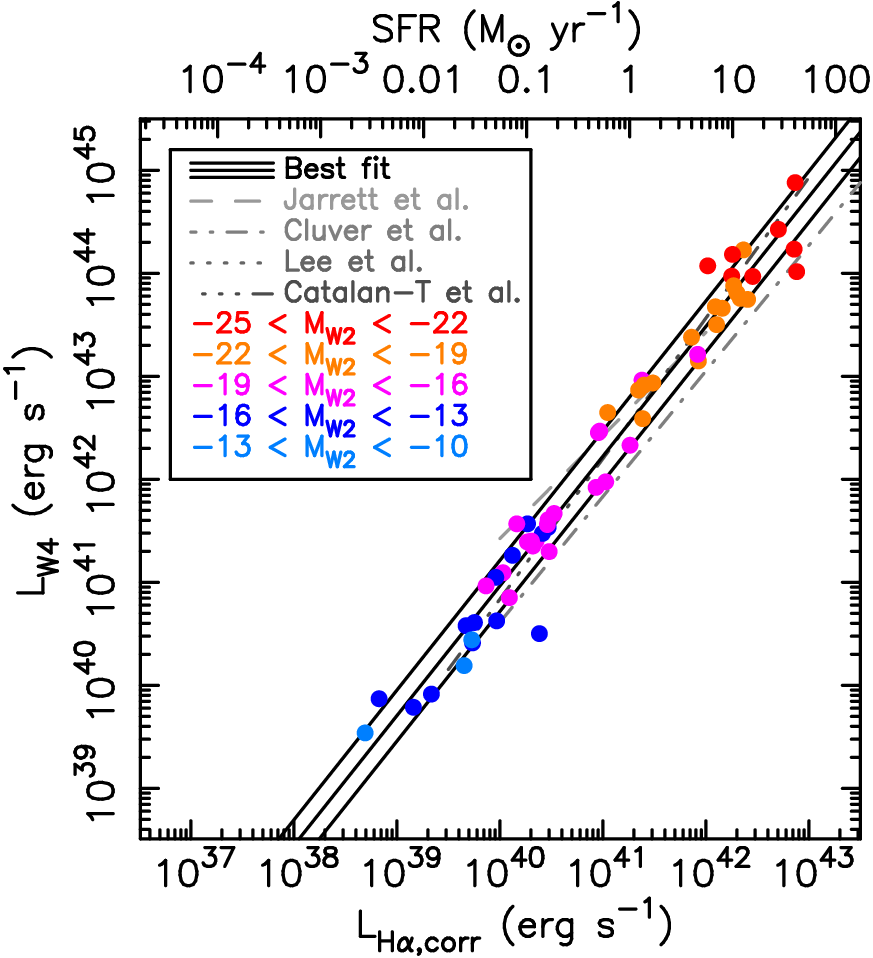}{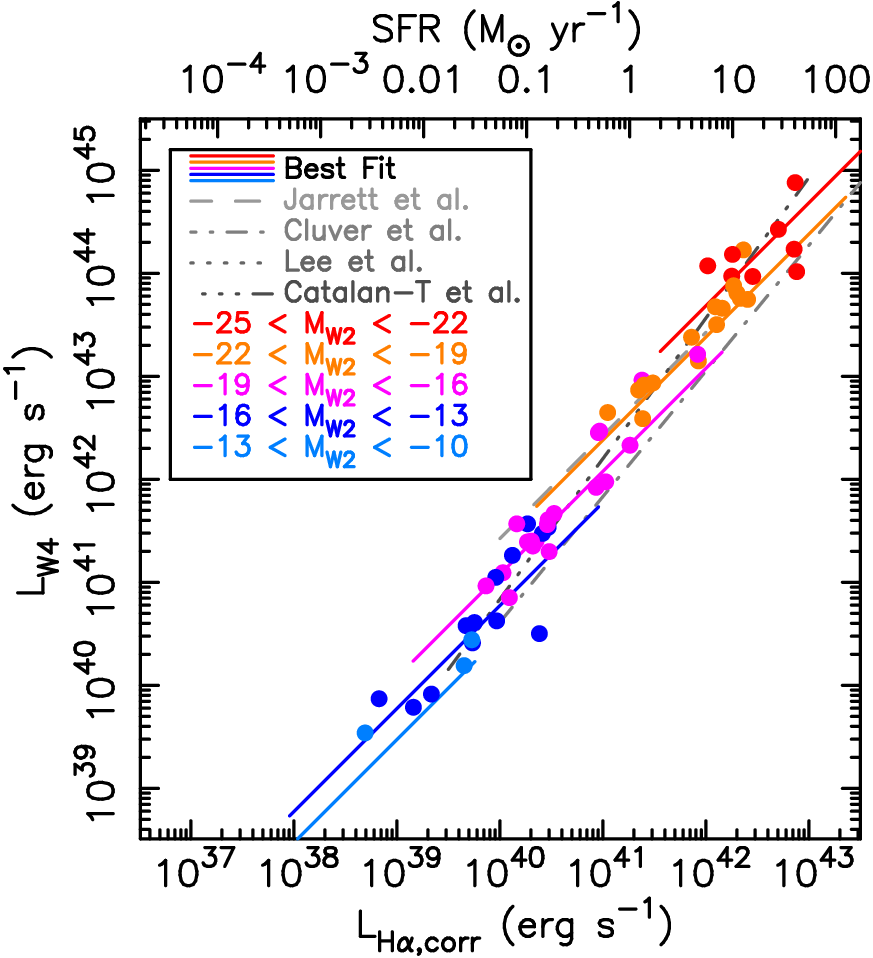}
\caption{WISE $W4$ luminosity as a function of Balmer decrement corrected H$\alpha$. While the index of the power-law fit (left panel) is comparable to power-law fits to {\it Spitzer} $8~{\rm \mu m}$ and WISE $W3$ ($12~{\rm \mu m}$) data, the scatter around the best-fit relation is significantly reduced.}
\label{fig:w4cal}
\end{figure*}

\begin{figure*}
\plottwo{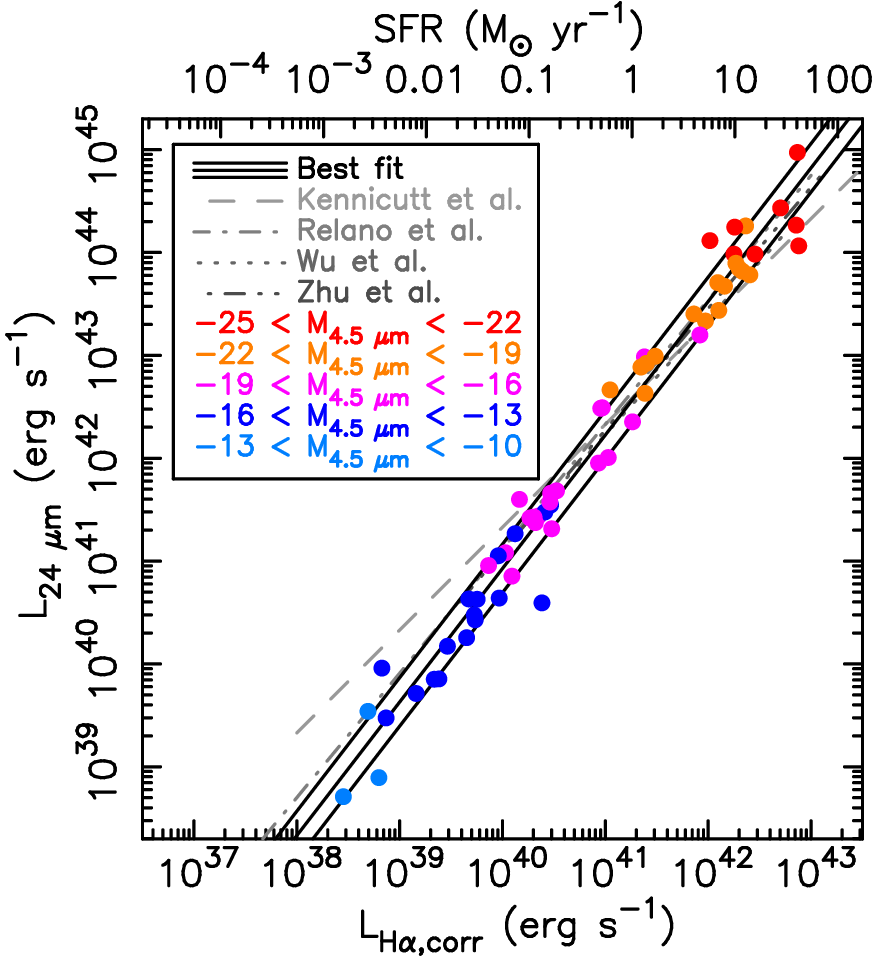}{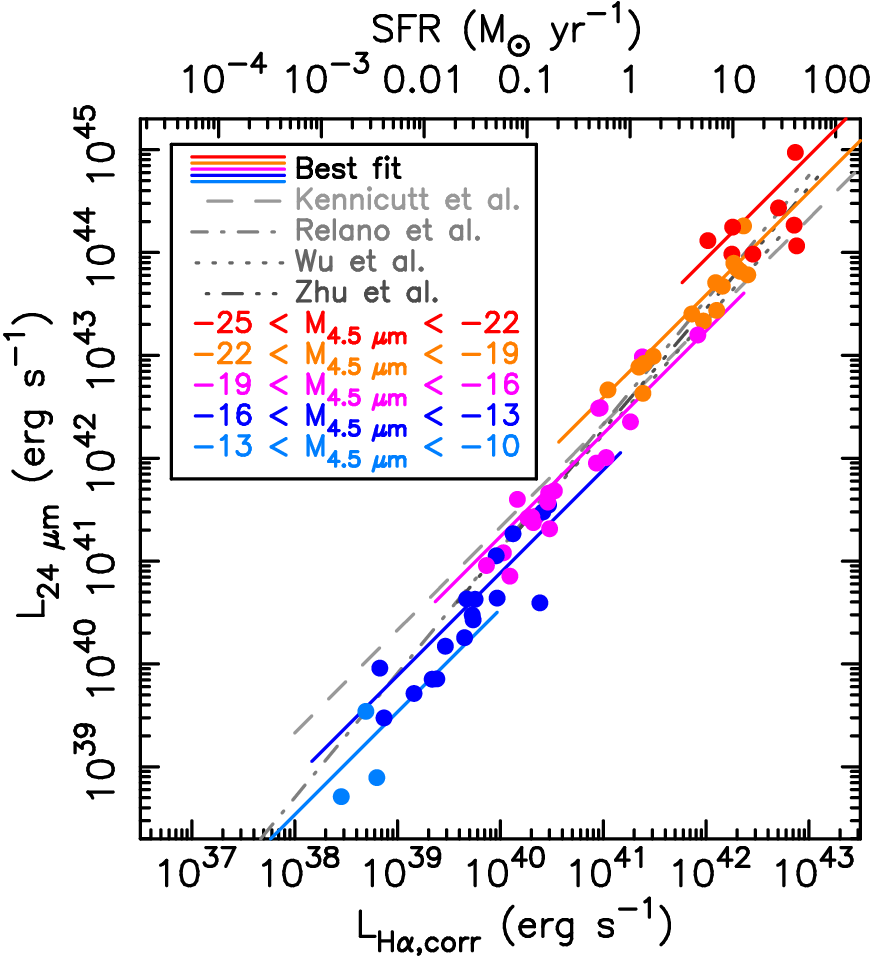}
\caption{Spitzer $24~{\rm \mu m}$ luminosity as a function of Balmer decrement corrected H$\alpha$, along with best-fit relations from the prior literature \citep{wu05,rel07,zhu08,ken09}. Compared to the relations for {\it Spitzer} $8~{\rm \mu m}$ and W3, there is better agreement between our calibration and those from the prior literature, although we still see offsets for the lowest luminosity galaxies.}
\label{fig:24cal}
\end{figure*}

We present the relationship between mid-infrared luminosity and Balmer decrement corrected H$\alpha$ luminosity in Figures~\ref{fig:I4cal} through to \ref{fig:24cal}. We have not subtracted stellar continuum from the mid-infrared luminosities (i.e., to produce a ``dust'' luminosity), as tests with the stellar continuum subtracted did not reduce the scatter and changed fit parameter values by $2\sigma$ or less. Figures~\ref{fig:I4cal}, ~\ref{fig:w3cal},  ~\ref{fig:w4cal} and ~\ref{fig:24cal} show the {\it Spitzer} IRAC $8~{\rm \mu m}$, WISE $W3$ ($12~{\rm \mu m}$), WISE $W4$ ($22.8~{\rm \mu m}$) and {\it Spitzer} MIPS $24~{\rm \mu m}$ respectively. In all of the figures grey lines denote power-law fits taken from a subset of the prior literature \citep{wu05,rel07,zhu08,ken09,jar13,lee13,clu14,cat15,dav16}. 

In Figures~\ref{fig:I4cal} through to \ref{fig:24cal} we provide power-law fits to the data and the relevant parameter values are provided in Table~\ref{table:calibration}. For all four mid-infrared bands we find power-law indices consistent with $1.3$. Some of the previous studies find or adopt power-law indices of close to unity \citep[i.e.,][]{cal07,ken09,jar13,lee13}, and when these fits are extrapolated to low luminosities they can disagree with our fits by an order of magnitude. However, given the mid-infrared emission from PAHs and dust depend on temperature and metallicity \citep[e.g.,][]{eng05,wu06,dra07,eng08,smi07,cal07}, there is no expectation that the power-law index for the mid-infrared calibrations for entire galaxies should be one.

Galaxies with H$\alpha$ luminosities of $10^{40}~{\rm erg~s^{-1}}$ have mid-infrared luminosities of $\sim 10^{40.8}~{\rm erg~s^{-1}}$ for all four mid-infrared bands. The scatter around the best-fit relations decreases with increasing wavelength, dropping from $0.33~{\rm dex}$ for {\it Spitzer} IRAC $8~{\rm \mu m}$ to $0.18~{\rm dex}$ for {\it Spitzer} MIPS $24~{\rm \mu m}$. The scatter is much larger than the uncertainties from the emission line measurements, photometry and distance errors, and we thus conclude the decreasing scatter with increasing wavelength is an intrinsic feature of these relations. 

Our fits to mid-infrared luminosity as a function of Balmer decrement corrected ${\rm H\alpha}$ luminosity (or star formation rates) have steeper power-law indices than those determined (or adopted) by the bulk of the prior literature \citep[the exception being][]{cat15}. Apart from when a power-law index of one is adopted \citep[e.g.,][]{ken09, jar13}, the largest discrepancies occur for studies that are limited to relatively high luminosities (i.e., $L_{H\alpha,Corr}>10^{40}~{\rm erg~s^{-1}}$). This includes most of the calibrations of {\it Spitzer} $8~{\rm \mu m}$ and WISE $W3$ from the prior literature. In contrast, studies that approach our luminosity limits, such as \citet{rel07} and \citet{cat15}, have power-laws indices that agree with ours to within 0.1. Furthermore, several previous studies show dwarf galaxies falling below their fits to the data \citep[e.g.,][]{wu05,ken09}. We thus conclude that differences between our power-law indices and those from the literature are primarily the result of our broad luminosity range, and that extrapolations of some relations from the prior literature can result in underestimates of SFRs.


In Figures~\ref{fig:I4cal} through to \ref{fig:24cal} the data-points are color coded by $\sim 4.5~{\rm \mu m}$ luminosity, which is a rough proxy for stellar mass. The luminosity-luminosity correlations present in the sample are clearly evident, and suggest the power-law fit parameters could depend on the mass range of the relevant calibration sample. Indeed, if we restrict our SFR calibrations to galaxies with $M_{\rm 4.5~\mu m}<-17$, the power-law indices for $8~{\rm \mu m}$ and $24~{\rm \mu m}$ relations decrease to $1.10 \pm 0.05$ and $1.19 \pm 0.05$ respectively, which is closer to values from some of the prior literature. The dependence of power-law indices on the stellar mass range of the sample flags a weakness of the power-law parameterization.

Our alternative to a power-law parameterization assumes SFR indicator luminosity scales linearly with SFR, with the normalization being a function of {\it Spitzer} $4.5~{\rm \mu m}$ or WISE $W2$ luminosity. Fits of this relation to the mid-infrared data are shown in the right hand panels of Figures~\ref{fig:I4cal} through to \ref{fig:24cal}, and fit parameters are presented in Table~\ref{table:calibration}. Effectively by construction, this parameterization agrees better with much of the literature for high mass galaxies, where the power-law indices (both measured and adopted) are close to one. However, the scatter of the data about the fits using this parameterization are (marginally) worse than the scatter of the data about the power-law fits. Thus, on the basis of the data presented in this paper alone, there is no compelling reason to use this parameterization in preference to a power-law, despite its potential aesthetic appeal.  

\subsection{Radio continuum}

We have determined radio continuum SFR calibrations at $1.4~{\rm GHz}$ and $150~{\rm MHz}$, which correspond to the frequencies of existing and planned wide-field radio continuum surveys from the Karl G. Jansky Very Large Array (JVLA), Low Frequency Array (LOFAR), Murchison Wide-field Array (MWA) and Australian Square Kilometre Array Pathfinder (ASAKP).  While the relationship between $150~{\rm MHz}$ luminosity and far-infrared luminosity has been studied previously \citep[e.g.,][]{cox88}, our work is one of the first direct calibrations of $150~{\rm MHz}$ as a SFR indicator \citep[e.g.,][G{\"u}rkan et al. in prep.]{cal17}. Radio continuum emission from star forming galaxies is dominated thermal bremsstrahlung and non-thermal synchrotron components. As bremsstrahlung and synchrotron are expected to have spectra with (roughly) $f_\nu \propto \nu^{-0.1}$ and $f_\nu \propto \nu^{-0.7}$ respectively, synchrotron should be increasingly dominant at longer wavelengths. Synchrotron is dominant at $1.4~{\rm GHz}$ in $\sim L^*$ galaxies, but synchrotron emission depends on cosmic ray production, magnetic field strength and galaxy size \citep[e.g.,][and references therein]{bel03}, so the bremsstrahlung component is increasingly important with decreasing galaxy mass. Consequently, we do not expect radio luminosity to be directly proportional to SFR.

\begin{figure*}
\plotone{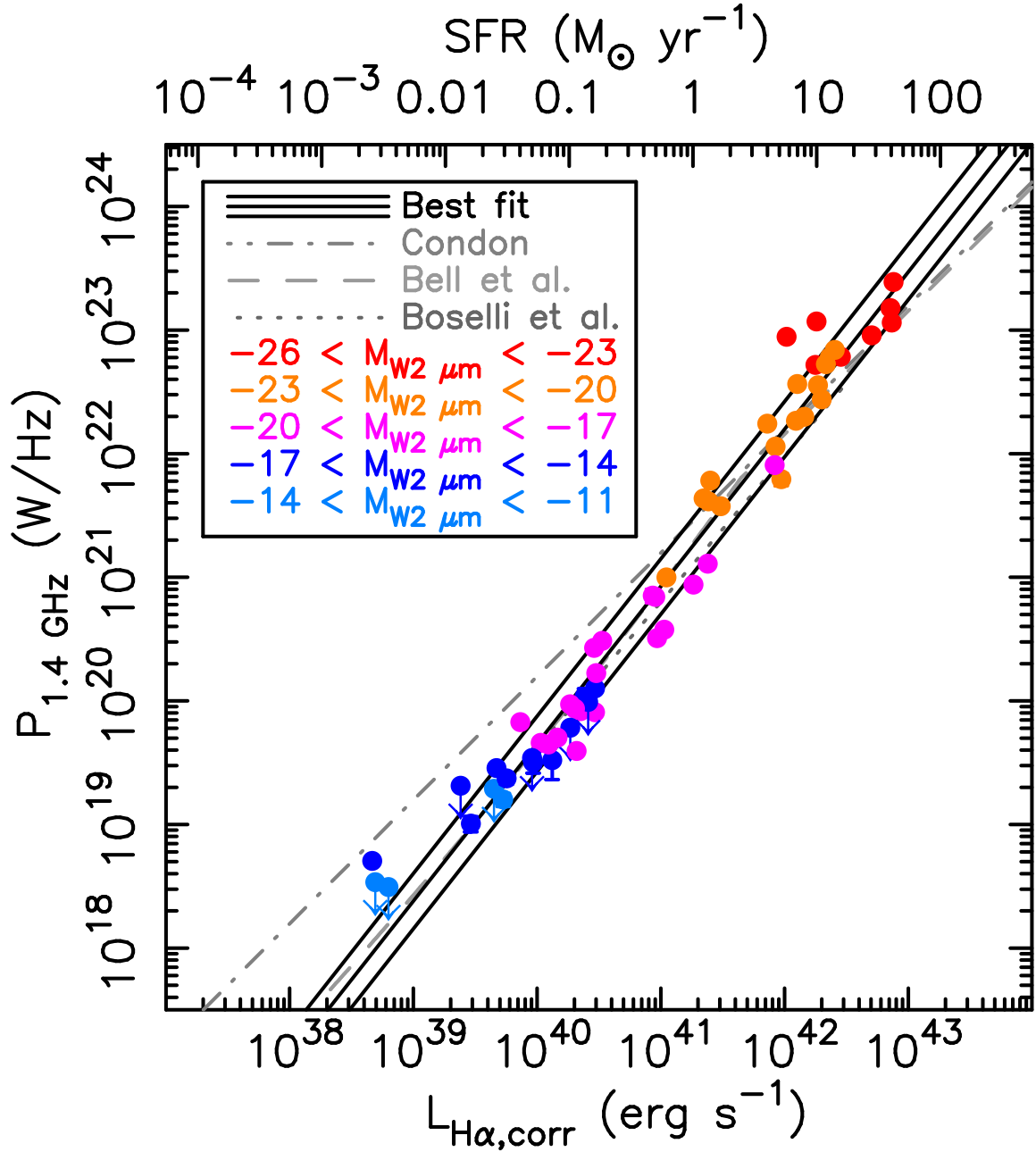}
\caption{$1.4~{\rm GHz}$ continuum luminosity as a function of Balmer decrement corrected ${\rm H\alpha}$, along with relations from the prior literature \citep{con92,bel03,bos15}. The scatter of the data around our best-fit power-law is less than $0.2~{\rm dex}$. At low radio luminosities we measure consistently higher ${\rm H\alpha}$ luminosities, and thus star formation rates, than the prior literature.}
\label{fig:radio14}
\end{figure*}

\begin{figure*}
\plotone{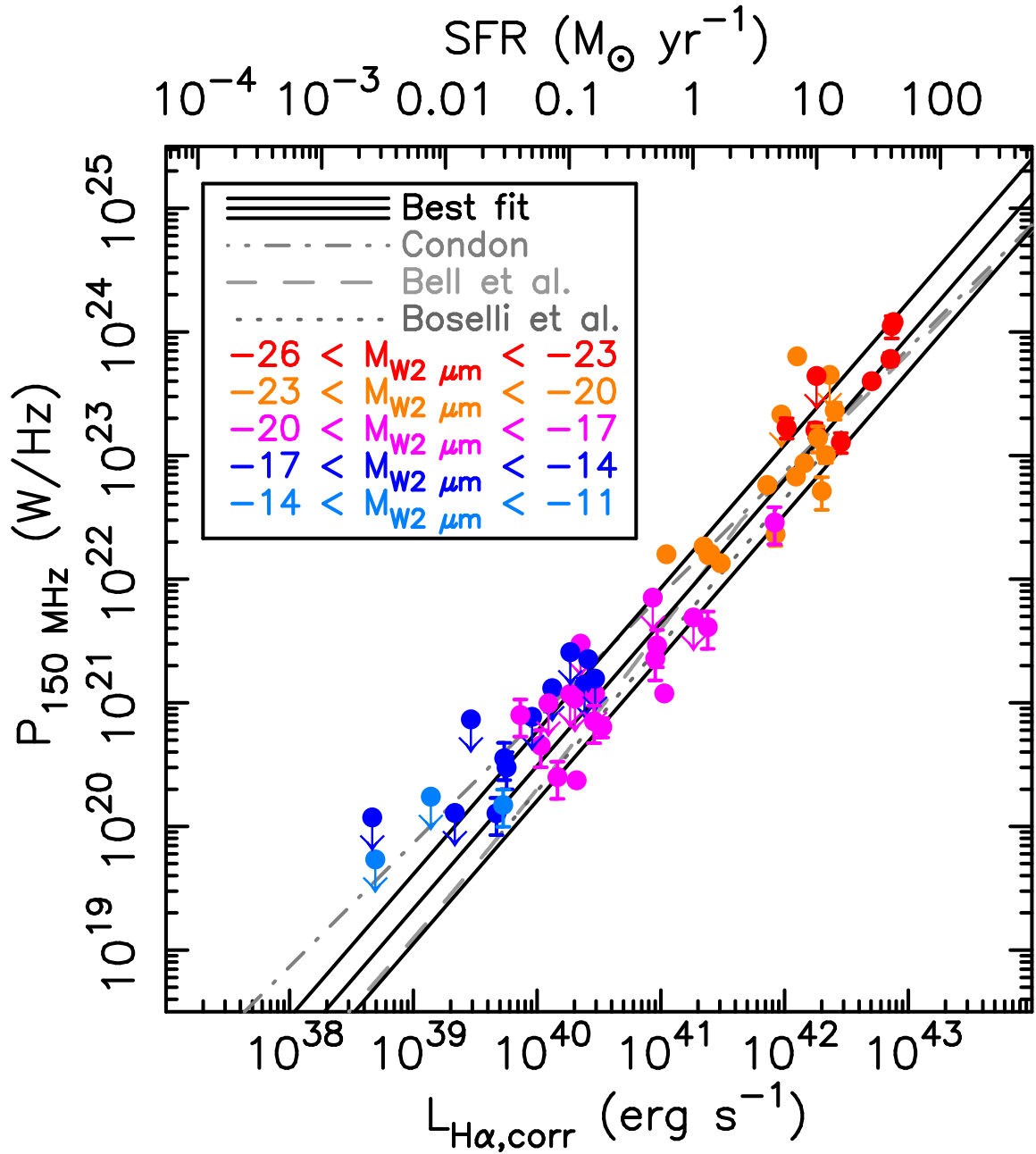}
\caption{$150~{\rm MHz}$ continuum luminosity as a function of Balmer decrement corrected ${\rm H\alpha}$. To plot relations from the prior literature \citep{con92,bel03,bos15}, we extrapolated radio luminosities from $1.4~{\rm GHz}$ to $150~{\rm MHz}$ by assuming $f_\nu \propto \nu^{-0.7}$. Despite changing an order of magnitude in wavelength, our best-fit power-law and the scatter of the data around this power-law are comparable to those measured at $1.4~{\rm GHz}$.}
\label{fig:radio15}
\end{figure*}

In Figures~\ref{fig:radio14} and~\ref{fig:radio15} we present the relationship between $1.4~{\rm GHz}$ and $150~{\rm MHz}$ (respectively) radio continuum power and Balmer decrement corrected H$\alpha$ luminosity. When fitting relations to the data, we only used radio sources with $>3\sigma$ flux density measurements, but in Figures~\ref{fig:radio14} and~\ref{fig:radio15} we also plot these upper limits. At $1.4~{\rm GHz}$, we find a power-law index of $1.27\pm 0.03$ and a scatter of just $0.18~{\rm dex}$, which is comparable to the $24~{\rm \mu m}$ calibration. At $150~{\rm MHz}$, we find a shallower power-law index of $1.16 \pm 0.05$ and a scatter of $0.24~{\rm dex}$. Our alternative parameterization (not plotted) performs no better than the power-law parameterization, with marginally worse scatter for both $1.4~{\rm GHz}$ and $150~{\rm MHz}$. 

Fits to the relationship between radio continuum luminosity and ${\rm H\alpha}$ luminosity from the prior literature are also plotted in Figures~\ref{fig:radio14} and~\ref{fig:radio15}. As there are no $150~{\rm MHz}$ versus ${\rm H\alpha}$ relations in the prior literature, we have extrapolated $1.4~{\rm GHz}$ calibrations to $150~{\rm MHz}$ by assuming $f_\nu \propto \nu^{-0.7}$. Relative to the mid-infrared relations, there is generally better agreement between power-law fits from the prior literature and our work. This agreement may result from the power-law indices of the radio continuum calibrations not being not being a strong function of the ${\rm H\alpha}$ luminosity and the sample stellar mass ranges. For example, when we restricted our calibrations to $M_{\rm 4.5~\mu m}<-17$ galaxies the power-law indices did not become significantly shallower. 

\section{Summary}
\label{sec:summary}

We have calibrated commonly used SFR indicators, including GALEX ultraviolet, {\it Spitzer} mid-infrared bands, WISE mid-infrared bands and radio continuum. This includes one of the first direct calibrations of $150~{\rm MHz}$ as a star formation rate indicator, which will be of use for new LOFAR and MWA wide-field surveys. The calibrations utilize 66 star forming galaxies, including galaxies drawn from the \citet{bro14} SED atlas and galaxies with distances less than $10~{\rm Mpc}$ with spectroscopy from \citet{mou06} and \citet{mou10}. Our sample includes a broad range of galaxy types, and has absolute magnitudes of $-24<M_r<-12$ and colors of $0.0<M_u-M_r<2.3$. The sample also spans five orders of magnitude in ${\rm H\alpha}$ luminosity, which is broader than much of the prior literature, and we thus provide improved calibrations of SFR indicators for dwarf galaxies. Systematic errors associated with aperture corrections have been mitigated by measuring ultraviolet and mid-infrared photometry with apertures matched to the same region as the spectrophotometry. To simplify transparency and reproducibility, all of the calibrations are anchored to Balmer decrement corrected H$\alpha$ luminosities, assuming 10,000 K Case B recombination and a \citet{fit99} dust attenuation curve. 

Our calibrations of SFR indicators are similar to those from the prior literature for $L^*$ galaxies, but for dwarf galaxies we often find that (for fixed broadband luminosity) SFRs are higher than what one would expect using (extrapolated) relations from the prior literature. We used two parameterizations to model the data, including the commonly used power-law relation and a linear relation where the normalization is a function of $4.5~{\rm \mu m}$ luminosity (a rough stellar mass proxy). We find the power-law parameterization provides better fits to the data, although there is no expectation that galaxies with the same SFR but different stellar masses and metallicities should have the same SFR indicator luminosity. Scatter of the data around best-fit relations is a function of wavelength, with the $1\sigma$ scatter being only $0.2~{\rm dex}$ for power-law fits to the WISE $W4$ ($22.8~{\rm \mu m}$), {\it Spitzer} $24~{\rm \mu m}$ and VLA $1.4$ GHz bands. We find $150~{\rm MHz}$ is only slightly worse than $1.4~{\rm GHz}$ as a star formation rate indicator, with the data having only $0.24~{\rm dex}$ scatter about the best-fit power-law for radio power as a function of ${\rm H\alpha}$ luminosity.

\begin{acknowledgments}


MJIB acknowledges financial support from The Australian Research Council (FT100100280), the Monash Research Accelerator Program (MRA), the Monash Outside Studies Programme (OSP) and the University of Cambridge. Part of this work was undertaken while MJIB was on OSP (sabbatical) leave at the University of Cambridge, Swinburne University and the University of Melbourne. Mederic Boquien was supported by MINEDUC-UA project, code ANT 1655.

This work is based in part on observations made with the {\it Spitzer} Space Telescope, obtained from the NASA/ IPAC Infrared Science Archive, both of which are operated by the Jet Propulsion Laboratory, California Institute of Technology under a contract with the National Aeronautics and Space Administration. This publication makes use of data products from the Wide-field Infrared Survey Explorer, which is a joint project of the University of California, Los Angeles, and the Jet Propulsion Laboratory/California Institute of Technology, funded by the National Aeronautics and Space Administration. We gratefully acknowledge NASA's support for construction, operation and science analysis for the GALEX mission, developed in cooperation with the Centre National d'Etudes Spatiales of France and the Korean Ministry of Science and Technology. 

The National Radio Astronomy Observatory is a facility of the National Science Foundation operated under cooperative agreement by Associated Universities, Inc. This research is based in part on observations taken with telescopes of the National Optical Astronomy Observatory, which is operated by the Association of Universities for Research in Astronomy (AURA) under cooperative agreement with the National Science Foundation.  We would like to thank the TIFR GMRT Sky Survey team for planning and carry out the original observations, and we offer a special thanks to the GMRT staff for their on-going support. GMRT is run by the National Centre for Radio Astrophysics of the Tata Institute of Fundamental Research. 

Funding for SDSS-III has been provided by the Alfred P. Sloan Foundation, the Participating Institutions, the National Science Foundation, and the U.S. Department of Energy Office of Science. The SDSS-III web site is http://www.sdss3.org/. SDSS-III is managed by the Astrophysical Research Consortium for the Participating Institutions of the SDSS-III Collaboration including the University of Arizona, the Brazilian Participation Group, Brookhaven National Laboratory, University of Cambridge, University of Florida, the French Participation Group, the German Participation Group, the Instituto de Astrofisica de Canarias, the Michigan State/Notre Dame/JINA Participation Group, Johns Hopkins University, Lawrence Berkeley National Laboratory, Max Planck Institute for Astrophysics, New Mexico State University, New York University, Ohio State University, Pennsylvania State University, University of Portsmouth, Princeton University, the Spanish Participation Group, University of Tokyo, University of Utah, Vanderbilt University, University of Virginia, University of Washington, and Yale University. The NASA-Sloan Atlas was created by Michael Blanton, with extensive help and testing from Eyal Kazin, Guangtun Zhu, Adrian Price-Whelan, John Moustakas, Demitri Muna, Renbin Yan and Benjamin Weaver. Funding for the NASA-Sloan Atlas has been provided by the NASA Astrophysics Data Analysis Program (08-ADP08-0072) and the NSF (AST-1211644).

{\it Facilities:} \facility{Bok (Boller \& Chivens spectrograph)}, \facility{CTIO:1.5m (R-C spectrograph)},  \facility{GMRT}, \facility{Sloan}, \facility{Spitzer (IRAC, IRS, MIPS)}, \facility{VLA}, \facility{WISE}

\end{acknowledgments}

\begin{deluxetable}{lll}
\tablecolumns{3}
\tablecaption{Ultraviolet and mid-infrared filter effective wavelengths.\label{table:filters}}
\tablehead{
  \colhead{Filter} &
  \colhead{Effective Wavelength} &
  \colhead{Reference}}
\startdata 
GALEX $FUV$                 & $ 1538.6$ ~  \AA   & \citet{mor07} \\
GALEX $NUV$                 & $ 2315.7$ ~  \AA   & \citet{mor07} \\
IRAC  $3.6~ {\rm \mu m}$  & $   3.55 ~ {\rm \mu m} $ & \citet{faz04} \\
IRAC  $4.5~ {\rm \mu m}$  & $  4.439 ~ {\rm \mu m} $ & \citet{faz04} \\
IRAC  $5.8~ {\rm \mu m}$  & $  5.731 ~ {\rm \mu m} $ & \citet{faz04} \\
IRAC  $8.0~ {\rm \mu m}$  & $  7.872 ~ {\rm \mu m} $ & \citet{faz04} \\
MIPS  $24~  {\rm \mu m}$  & $ 23.675 ~ {\rm \mu m} $ & \citet{eng07} \\
WISE  $W1$                  & $ 3.3526 ~ {\rm \mu m} $ & \citet{jar11} \\
WISE  $W2$                  & $ 4.6028 ~ {\rm \mu m} $ & \citet{jar11} \\
WISE  $W3$                  & $ 11.5608 ~ {\rm \mu m} $ & \citet{jar11} \\
WISE  $W4$                  & $   22.8 ~ {\rm \mu m} $ & \citet{bro14b} \\

\\
\hline
\enddata
\end{deluxetable}


\begin{longrotatetable}
\begin{deluxetable*}{lcccccccccccc}
\tablecolumns{13}
\tabletypesize{\tiny}
\tablecaption{Summary of galaxy properties, including aperture emission line fluxes and (total) radio continuum flux densities.\label{table:eflux}}
\tablehead{
  \colhead{Name} &
  \colhead{$d_L$} &
  \colhead{a} &
  \colhead{b} &
  \colhead{P.A.} &
  \colhead{$m_{g,total}$}  &
  \colhead{$m_{g,aper}$} &
 \colhead{H$\beta$ $\lambda 4861$} &
 \colhead{O[III] $\lambda 5007$} &
 \colhead{H$\alpha$ $\lambda 6563$} &
 \colhead{[NII] $\lambda 6716$} &
 \colhead{1.4 GHz} &
 \colhead{150 MHz} \\
  \colhead{} &
  \colhead{(Mpc)} &
  \colhead{($^{\prime\prime}$)} &
  \colhead{($^{\prime\prime}$)} &
  \colhead{($^\circ$)} &
  \colhead{}  &
 \colhead{}  &
\multicolumn{4}{c}{($10^{-14}~{\rm erg~cm^{-2}~s^{-1}}$)} &
 \colhead{(mJy)} &
 \colhead{(mJy)}  
}
\tabletypesize{\scriptsize}
\startdata 
    Arp 256 N        &  110.3 &  40 &  60 &  90 & 14.32 & 14.32 & $12.9 \pm  0.6 $ & $10.6 \pm  0.8 $ & $44.5 \pm  1.9 $ & $12.9 \pm  1.4 $ & $   4 $& $  23 $ \\ 
    Arp 256 S        &  109.4 &  40 &  40 &  90 & 14.36 & 14.36 & $15.4 \pm  0.6 $ & $14.0 \pm  0.5 $ & $68.8 \pm  1.4 $ & $23.1 \pm  0.9 $ & $  42 $& $ 158 $ \\ 
    NGC 0337         &   18.0 &  95 &  55 &  70 & 11.48 & 11.98 & $73.4 \pm  1.7 $ & $101.1 \pm  1.5 $ & $ 261 \pm    3 $ & $48.2 \pm  2.0 $ & $ 106 $& $ 404 $\tablenotemark{b} \\ 
    CGCG 436-030     &  125.1 &  35 &  40 &  90 & 14.58 & 14.58 & $ 6.6 \pm  0.5 $ & $ 5.7 \pm  0.5 $ & $32.1 \pm  1.1 $ & $13.6 \pm  1.0 $ & $  50 $& $  87 $ \\ 
    NGC 0520         &   30.5 & 140 & 100 &  90 & 11.98 & 11.98 & $14.9 \pm  2.8 $ & $12.0 \pm  2.3 $ & $41.0 \pm  6.5 $ & $26.8 \pm  4.5 $ & $ 176 $& $ 433 $\tablenotemark{b} \\ 
         NGC 0628    &   10.1 & 346 &  55 &  70 &  9.27 & 10.92 & $37.9 \pm  5.9 $ &       $ < 13.2 $ & $103.4 \pm  8.1 $ & $38.2 \pm  6.7 $ & $ 180 $& $ 321 $ \\ 
    III Zw 035       &  109.5 &  20 &  35 &  90 & 15.17 & 15.17 & $ 1.7 \pm  0.2 $ & $ 3.3 \pm  0.2 $ & $ 8.5 \pm  0.5 $ & $ 5.1 \pm  0.4 $ & $  40 $& $  56 $ \\ 
    NGC 0695         &  130.2 &  60 &  45 &  90 & 13.50 & 13.50 & $23.4 \pm  0.7 $ & $12.8 \pm  0.6 $ & $ 118 \pm    2 $ & $42.6 \pm  1.0 $ & $  75 $& $ 301 $\tablenotemark{b} \\ 
         NGC 0855    &    8.8 &  86 &  55 &  70 & 12.52 & 12.81 & $23.8 \pm  1.2 $ & $37.4 \pm  1.3 $ & $74.4 \pm  2.2 $ & $ 9.2 \pm  2.0 $ & $   5 $& $  48 $ \\ 
    NGC 1144         &  115.3 &  50 &  60 &  90 & 13.30 & 13.30 & $15.3 \pm  1.1 $ & $15.0 \pm  1.2 $ & $99.7 \pm  4.1 $ & $13.3 \pm  2.4 $ & $ 155 $& $ 763 $\tablenotemark{b} \\ 
    NGC 1275         &   62.5 &  75 &  40 &  90 & 11.21 & 12.00 & $46.4 \pm  2.8 $ & $104.3 \pm  2.9 $ & $ 185 \pm    7 $ & $ 153 \pm    7 $ & $ 22.8 \times 10^3 $& $ 55.6 \times 10^3 $\tablenotemark{a} \\ 
    NGC 1614         &   64.2 &  80 &  60 &  90 & 12.49 & 12.77 & $35.4 \pm  0.9 $ & $25.5 \pm  0.8 $ & $ 193 \pm    2 $ & $ 103 \pm    2 $ & $ 137 $& $ 340 $\tablenotemark{b} \\ 
    NGC 2388         &   60.3 &  60 &  30 &  90 & 13.62 & 13.86 & $ 4.4 \pm  0.4 $ & $ 2.5 \pm  0.4 $ & $38.5 \pm  1.0 $ & $19.9 \pm  0.6 $ & $  75 $& $ 215 $\tablenotemark{a} \\ 
         NGC 2403    &    3.2 & 657 &  56 & 127 &  8.18 &  9.71 & $ 417 \pm   40 $ & $ 357 \pm   41 $ & $1484 \pm   74 $ & $ 322 \pm   52 $ & $ 330 $& $ 304 $ \\ 
    NGC 2537         &    8.1 & 100 &  60 &  90 & 11.98 & 11.98 & $75.0 \pm  1.9 $ & $114.6 \pm  1.6 $ & $ 237 \pm    4 $ & $42.9 \pm  2.7 $ & $  10 $& $ 191 $ \\ 
    UGC 04881        &  164.3 &  60 &  40 &  90 & 14.30 & 14.30 & $ 2.9 \pm  0.4 $ & $ 1.3 \pm  0.3 $ & $15.8 \pm  1.0 $ & $ 7.5 \pm  0.7 $ & $  37 $& $  69 $ \\ 
         NGC 2798    &   28.6 &  84 &  55 & 103 & 12.28 & 12.81 & $25.9 \pm  1.2 $ & $12.4 \pm  1.1 $ & $ 125 \pm    2 $ & $62.0 \pm  1.9 $ & $  82 $& $ 298 $\tablenotemark{a} \\ 
    UGCA 166         &   19.0 &  20 &  20 &  90 & 15.76 & 15.83 & $19.5 \pm  0.2 $ & $33.4 \pm  0.2 $ & $54.6 \pm  0.5 $ & $ 1.2 \pm  0.4 $ & $   2 $& $  11 $ \\ 
    NGC 3049         &   17.2 &  72 &  55 & 115 & 12.72 & 13.29 & $25.0 \pm  0.8 $ & $ 9.6 \pm  0.7 $ & $93.8 \pm  1.7 $ & $32.8 \pm  1.2 $ & $   9 $& $  82 $ \\ 
    NGC 3079         &   20.6 & 100 & 330 &  90 & 11.03 & 11.03 & $95.1 \pm 11.8 $ & $115.7 \pm 10.0 $ & $ 420 \pm   29 $ & $ 161 \pm   18 $ & $ 865 $& $ 4.0 \times 10^3 $\tablenotemark{a} \\ 
    UGCA 208         &   28.5 &  30 &  20 &  90 & 14.81 & 15.03 & $ 9.6 \pm  0.3 $ & $25.1 \pm  0.3 $ & $ 6.4 \pm  0.6 $ & $ 3.9 \pm  0.4 $ & $   1 $& $  38 $ \\ 
    NGC 3198         &   12.9 & 180 &  55 & 120 & 10.80 & 12.18 & $18.7 \pm  1.5 $ & $19.2 \pm  1.8 $ & $86.4 \pm  2.8 $ & $28.1 \pm  2.3 $ & $  38 $& $ 219 $ \\ 
    NGC 3265         &   24.0 &  42 &  55 & 120 & 13.55 & 13.76 & $ 8.2 \pm  0.6 $ & $ 4.8 \pm  0.5 $ & $40.0 \pm  1.1 $ & $16.1 \pm  0.8 $ & $  10 $& $  33 $ \\ 
         Mrk 33      &   25.5 &  33 &  55 & 110 & 13.16 & 13.30 & $52.9 \pm  0.7 $ & $92.7 \pm  0.7 $ & $ 185 \pm    1 $ & $34.7 \pm  0.9 $ & $  17 $& $  53 $ \\ 
    NGC 3310         &   19.2 &  90 &  65 &  90 & 11.05 & 11.05 & $ 369 \pm    3 $ & $ 704 \pm    3 $ & $1231 \pm    6 $ & $ 272 \pm    4 $ & $ 397 $& $ 1.3 \times 10^3 $\tablenotemark{a} \\ 
    NGC 3351         &   10.5 & 245 &  55 & 115 & 10.05 & 10.90 & $60.1 \pm  3.8 $ & $25.2 \pm  4.6 $ & $ 229 \pm    6 $ & $101.5 \pm  5.5 $ & $  43 $& $ 623 $ \\ 
    UGCA 219         &   38.6 &  35 &  30 &  90 & 14.47 & 14.72 & $14.9 \pm  0.3 $ & $49.8 \pm  0.4 $ & $40.9 \pm  0.8 $ & $ 2.8 \pm  0.8 $ & $   4 $& $  20 $ \\ 
    NGC 3521         &   13.8 & 263 &  56 & 110 &  9.18 & 10.18 & $ 107 \pm    8 $ & $49.1 \pm  7.3 $ & $ 451 \pm   13 $ & $ 189 \pm   10 $ & $ 375 $& $ 4.3 \times 10^3 $ \\ 
         NGC 3627    &    9.0 & 200 &  55 & 115 &  9.21 & 10.22 & $74.4 \pm  5.1 $ & $49.1 \pm  5.9 $ & $ 330 \pm   12 $ & $ 147 \pm    9 $ & $ 453 $& $ 1.5 \times 10^3 $\tablenotemark{b} \\ 
    IC 0691          &   22.7 &  40 &  40 &  90 & 13.90 & 14.08 & $27.5 \pm  0.6 $ & $52.0 \pm  0.7 $ & $99.1 \pm  1.2 $ & $24.4 \pm  0.9 $ & $  16 $& $ 258 $\tablenotemark{a} \\ 
    NGC 3690         &   48.5 &  90 &  60 &  90 & 12.03 & 12.03 & $ 163 \pm    2 $ & $ 176 \pm    2 $ & $ 708 \pm    5 $ & $ 279 \pm    3 $ & $ 677 $& $ 4.6 \times 10^3 $\tablenotemark{a} \\ 
    NGC 3773         &   10.8 &  38 &  55 & 115 & 12.80 & 13.47 & $30.8 \pm  0.6 $ & $39.7 \pm  0.6 $ & $96.1 \pm  1.2 $ & $16.1 \pm  0.9 $ & $   6 $& $  42 $ \\ 
    Mrk 1450         &   19.0 &  20 &  15 &  90 & 15.34 & 15.34 & $19.6 \pm  0.2 $ & $103.3 \pm  0.3 $ & $57.3 \pm  0.5 $ & $ 2.3 \pm  0.3 $ & $ 0.9 $& $  20 $ \\ 
    UGC 06665        &   84.2 &  30 &  60 &  90 & 13.85 & 13.85 & $39.4 \pm  1.1 $ & $103.0 \pm  1.3 $ & $ 145 \pm    2 $ & $23.6 \pm  1.5 $ & $  33 $& $  61 $ \\ 
    NGC 3870         &   11.8 &  50 &  40 &  90 & 13.10 & 13.37 & $20.9 \pm  0.8 $ & $34.0 \pm  0.9 $ & $70.1 \pm  1.6 $ & $12.5 \pm  1.0 $ & $   5 $& $  32 $ \\ 
    UM 461           &   12.7 &  25 &  20 &  90 & 15.35 & 15.71 & $11.2 \pm  0.2 $ & $63.6 \pm  0.3 $ & $32.7 \pm  0.5 $ & $ 0.7 \pm  0.3 $ & $ 0.4 $& $   6 $ \\ 
    UGC 06850        &   13.5 &  36 &  40 &  90 & 14.14 & 14.18 & $43.4 \pm  0.7 $ & $ 171 \pm    1 $ & $ 126 \pm    1 $ & $ 4.6 \pm  1.0 $ & $   6 $& $  36 $ \\ 
    NGC 4088         &   12.8 & 140 & 300 & 135 & 10.79 & 10.79 & $ 138 \pm    9 $ & $68.7 \pm  8.7 $ & $ 569 \pm   18 $ & $ 192 \pm   14 $ & $ 222 $& $ 940 $\tablenotemark{a} \\ 
    NGC 4138         &   16.0 &  60 & 120 &  90 & 11.54 & 11.81 & $23.7 \pm  2.1 $ & $25.6 \pm  2.1 $ & $71.2 \pm  4.2 $ & $42.8 \pm  2.9 $ & $  19 $& $  43 $ \\ 
    NGC 4194         &   40.8 & 125 &  30 & 165 & 12.82 & 12.82 & $45.3 \pm  0.6 $ & $42.7 \pm  0.6 $ & $ 204 \pm    2 $ & $94.6 \pm  1.1 $ & $ 101 $& $ 289 $\tablenotemark{a} \\ 
    Haro 06          &   13.7 &  30 &  20 &  90 & 14.77 & 14.77 & $19.4 \pm  0.3 $ & $60.7 \pm  0.3 $ & $56.5 \pm  0.5 $ & $ 4.7 \pm  0.4 $ & $   1 $& $   9 $ \\ 
    NGC 4254         &   13.9 & 177 &  55 & 120 &  9.94 & 10.91 & $116.4 \pm  4.2 $ & $24.9 \pm  4.0 $ & $ 611 \pm   11 $ & $ 181 \pm    8 $ & $ 420 $& $ 2.3 \times 10^3 $\tablenotemark{b} \\ 
    NGC 4321         &   13.9 & 245 &  56 & 121 & 10.32 & 10.82 & $102.7 \pm  5.3 $ & $30.3 \pm  5.8 $ & $ 394 \pm   12 $ & $ 149 \pm    7 $ & $ 263 $& $ 700 $ \\ 
    NGC 4385         &   34.5 & 100 &  60 &  90 & 11.94 & 12.83 & $36.4 \pm  2.3 $ & $21.8 \pm  2.2 $ & $122.7 \pm  4.2 $ & $57.2 \pm  2.9 $ & $  13 $& $  62 $ \\ 
         NGC 4536    &   12.4 & 250 &  55 & 115 & 11.17 & 11.44 & $49.1 \pm  3.1 $ & $16.9 \pm  3.2 $ & $ 304 \pm    8 $ & $134.0 \pm  5.7 $ & $ 205 $& $ 732 $\tablenotemark{b} \\ 
         NGC 4559    &    7.3 & 354 &  55 & 135 & 10.29 & 10.92 & $ 153 \pm    5 $ & $ 154 \pm    5 $ & $ 571 \pm   11 $ & $132.8 \pm  7.3 $ & $  59 $& $ 187 $\tablenotemark{a} \\ 
         NGC 4579    &   16.7 & 194 &  55 & 120 & 10.46 & 10.82 & $20.5 \pm  4.0 $ & $40.3 \pm  3.6 $ & $84.1 \pm  9.3 $ & $107.1 \pm  6.3 $ & $  97 $& $ 646 $\tablenotemark{b} \\ 
    NGC 4625         &   10.5 &  72 &  56 & 143 & 12.52 & 12.86 & $15.8 \pm  1.5 $ & $ 4.4 \pm  1.3 $ & $60.2 \pm  2.7 $ & $25.1 \pm  2.2 $ & $   7 $& $  44 $ \\ 
         NGC 4631    &    7.3 & 512 &  56 & 101 &  9.45 & 10.41 & $ 244 \pm    7 $ & $ 361 \pm    7 $ & $ 957 \pm   14 $ & $ 206 \pm   12 $ & $ 982 $& $ 3.8 \times 10^3 $\tablenotemark{a} \\ 
    NGC 4670         &   23.1 &  80 &  40 &  90 & 12.71 & 12.81 & $75.1 \pm  1.5 $ & $ 174 \pm    1 $ & $ 230 \pm    2 $ & $28.1 \pm  1.5 $ & $  14 $& $  38 $ \\ 
         NGC 4826    &    5.2 & 330 &  55 & 120 &  8.93 &  9.58 & $102.2 \pm 10.9 $ & $93.4 \pm 11.0 $ & $ 438 \pm   23 $ & $ 271 \pm   18 $ & $ 103 $& $ 208 $ \\ 
         NGC 5033    &   19.3 & 354 &  56 & 150 & 11.15 & 11.21 & $39.8 \pm  3.4 $ & $33.0 \pm  3.8 $ & $ 237 \pm   10 $ & $119.8 \pm  7.6 $ & $ 205 $& $ 971 $\tablenotemark{a} \\ 
    IC 0860          &   53.8 &  30 &  40 &  90 & 14.00 & 14.05 & $ 2.7 \pm  0.4 $ &       $ <  0.7 $ & $ 2.6 \pm  0.5 $ &       $ <  0.4 $ & $  31 $& $  59 $ \\ 
    UGC 08335 SE     &  132.5 &  35 &  50 &  90 & 14.95 & 14.95 & $ 7.5 \pm  0.7 $ & $ 9.1 \pm  0.6 $ & $42.4 \pm  1.6 $ & $18.6 \pm  1.1 $ & $  57 $& $ 112 $ \\ 
    NGC 5194         &    7.3 & 370 &  56 & 158 &  9.88 &  9.88 & $ 158 \pm   14 $ & $38.6 \pm 15.2 $ & $ 819 \pm   34 $ & $ 451 \pm   23 $ & $ 1.6 \times 10^3 $& $ 6.5 \times 10^3 $\tablenotemark{a} \\ 
    NGC 5256         &  120.9 &  50 &  30 &  90 & 13.48 & 13.75 & $23.6 \pm  0.6 $ & $97.8 \pm  0.8 $ & $91.9 \pm  2.5 $ & $50.1 \pm  1.2 $ & $ 126 $& $ 607 $\tablenotemark{a} \\ 
    NGC 5257         &  102.4 &  70 &  90 &  90 & 13.11 & 13.11 & $32.6 \pm  1.9 $ & $22.7 \pm  1.6 $ & $ 126 \pm    4 $ & $44.3 \pm  3.4 $ & $  49 $& $ 103 $ \\ 
    NGC 5258         &  101.8 &  80 &  90 &  90 & 13.09 & 13.09 & $14.2 \pm  1.9 $ & $ 4.7 \pm  1.5 $ & $62.5 \pm  4.3 $ & $24.7 \pm  2.8 $ & $  42 $& $ 130 $ \\ 
    UGC 08696        &  160.5 &  15 &  85 &  90 & 14.67 & 14.67 & $ 4.9 \pm  0.4 $ & $26.1 \pm  0.3 $ & $23.9 \pm  1.4 $ & $27.9 \pm  1.1 $ & $ 163 $& $ 337 $\tablenotemark{a} \\ 
    NGC 5653         &   48.1 &  80 &  30 &  90 & 12.52 & 12.73 & $31.3 \pm  0.8 $ & $ 7.1 \pm  0.8 $ & $ 154 \pm    2 $ & $59.0 \pm  1.1 $ & $  72 $& $ 314 $\tablenotemark{a} \\ 
    Mrk 0475         &    9.2 &  30 &  20 &  90 & 15.49 & 15.82 & $11.9 \pm  0.2 $ & $59.2 \pm  0.2 $ & $33.5 \pm  0.4 $ & $ 1.0 \pm  0.3 $ & $ 0.6 $& -     \\ 
         NGC 5713    &   31.3 &  90 &  55 & 113 & 11.30 & 11.76 & $59.2 \pm  1.5 $ & $22.0 \pm  1.5 $ & $ 276 \pm    3 $ & $110.9 \pm  2.3 $ & $ 158 $& $ 578 $\tablenotemark{b} \\ 
    UGC 09618 S      &  145.5 &  30 &  47 &  90 & 14.49 & 14.49 & $ 6.2 \pm  0.4 $ & $ 4.1 \pm  0.4 $ & $25.2 \pm  0.8 $ & $ 7.9 \pm  0.6 $ & $  15 $& $ 253 $ \\ 
    UGC 09618        &  143.1 &  30 & 110 &  90 & 13.85 & 13.85 & $10.6 \pm  0.7 $ & $ 9.7 \pm  0.6 $ & $31.4 \pm  1.5 $ & $26.2 \pm  1.4 $ & $  82 $& $ 454 $ \\ 
    UGC 09618 N      &  146.4 &  30 &  50 &  90 & 14.68 & 14.68 & $ 4.7 \pm  0.4 $ & $ 6.0 \pm  0.5 $ & $17.9 \pm  1.9 $ & $13.0 \pm  1.1 $ & $  82 $& $ 454 $ \\ 
    NGC 5953         &   34.3 &  60 &  75 &  90 & 12.49 & 12.49 & $31.8 \pm  1.7 $ & $30.5 \pm  1.6 $ & $ 151 \pm    4 $ & $86.8 \pm  2.7 $ & $  75 $& $ 276 $\tablenotemark{b} \\ 
    UGCA 410         &   15.4 &  25 &  15 &  90 & 14.95 & 15.12 & $15.4 \pm  0.1 $ & $80.9 \pm  0.2 $ & $47.8 \pm  0.4 $ & $ 2.4 \pm  0.2 $ & $ 0.8 $& $   4 $ \\ 
    NGC 5992         &  137.2 &  40 &  40 &  90 & 13.84 & 14.01 & $11.1 \pm  0.4 $ & $ 9.4 \pm  0.4 $ & $41.4 \pm  0.9 $ & $12.7 \pm  0.7 $ & $  16 $& $  62 $ \\ 
    NGC 6052         &   73.3 &  50 &  60 &  90 & 12.96 & 12.98 & $66.3 \pm  1.2 $ & $ 118 \pm    1 $ & $ 242 \pm    3 $ & $40.7 \pm  1.8 $ & $ 108 $& $ 362 $\tablenotemark{b} \\ 
    NGC 6090         &  126.1 &  45 &  20 &  90 & 13.89 & 14.11 & $21.9 \pm  0.4 $ & $14.0 \pm  0.3 $ & $95.6 \pm  0.9 $ & $39.1 \pm  0.6 $ & $  48 $& $ 212 $\tablenotemark{a} \\ 
    NGC 6240         &  108.8 &  50 &  80 &  90 & 13.07 & 13.07 & $18.0 \pm  1.6 $ & $27.0 \pm  1.7 $ & $81.2 \pm 14.2 $ & $ 126 \pm   16 $ & $ 426 $& $ 2.4 \times 10^3 $\tablenotemark{b} \\ 
    II Zw 096        &  150.0 &  50 &  40 &  90 & 13.49 & 13.91 & $24.5 \pm  0.6 $ & $41.9 \pm  0.6 $ & $97.3 \pm  1.7 $ & $27.4 \pm  1.3 $ & $  43 $& $ 418 $\tablenotemark{b} \\ 
         NGC 7331    &   13.9 & 200 &  55 &  90 &  9.58 & 10.48 & $59.2 \pm  6.6 $ & $59.2 \pm  6.6 $ & $ 240 \pm   12 $ & $98.0 \pm  9.9 $ & $ 329 $& $ 1.5 \times 10^3 $ \\ 
    CGCG 453-062     &  102.3 &  50 &  30 &  90 & 14.02 & 14.02 & $ 8.0 \pm  0.6 $ & $ 3.5 \pm  0.6 $ & $45.9 \pm  1.4 $ & $21.3 \pm  1.1 $ & $  43 $& $  81 $ \\ 
    IC 5298          &  111.6 &  40 &  30 &  90 & 14.11 & 14.31 & $ 3.6 \pm  0.3 $ & $ 5.3 \pm  0.3 $ & $18.1 \pm  0.8 $ & $14.8 \pm  0.6 $ & $  35 $& $ 103 $ \\ 
    NGC 7591         &   53.2 &  80 &  75 &  90 & 13.26 & 13.26 & $ 8.9 \pm  1.1 $ & $ 5.7 \pm  1.2 $ & $41.7 \pm  2.0 $ & $21.7 \pm  1.6 $ & $  52 $& $ 184 $ \\ 
    NGC 7592         &   99.5 &  60 &  58 &  90 & 13.55 & 13.55 & $12.9 \pm  0.7 $ & $ 9.3 \pm  0.5 $ & $49.5 \pm  1.3 $ & $17.1 \pm  1.0 $ & $  75 $& $ 143 $\tablenotemark{b} \\ 
    NGC 7673         &   47.0 &  70 &  40 &  90 & 12.89 & 12.89 & $63.1 \pm  0.7 $ & $108.8 \pm  0.7 $ & $ 220 \pm    2 $ & $35.1 \pm  1.2 $ & $  43 $& $  87 $ \\ 
    NGC 7679         &   57.3 &  80 &  30 &  90 & 12.53 & 12.79 & $44.0 \pm  1.0 $ & $55.5 \pm  1.2 $ & $ 179 \pm    2 $ & $72.1 \pm  1.5 $ & $  56 $& $ 135 $ \\ 
    Mrk 0930         &   74.6 &  25 &  15 &  90 & 14.57 & 14.65 & $32.0 \pm  0.4 $ & $ 133 \pm    0 $ & $98.7 \pm  0.7 $ & $ 5.6 \pm  0.4 $ & $  12 $& $  43 $ \\ 
    NGC 7714         &   38.5 &  60 &  60 &  90 & 12.44 & 12.44 & $118.2 \pm  2.0 $ & $ 201 \pm    2 $ & $ 434 \pm    5 $ & $ 144 \pm    3 $ & $  66 $& $ 119 $ \\ 
    NGC 7771         &   58.0 & 130 &  50 &  90 & 12.25 & 12.25 & $14.8 \pm  1.4 $ & $ 8.8 \pm  1.5 $ & $103.5 \pm  4.9 $ & $46.5 \pm  3.2 $ & $ 141 $& $ 458 $ \\ 
    Mrk 0331         &   74.9 &  40 &  40 &  90 & 13.70 & 13.83 & $ 9.1 \pm  0.5 $ & $ 4.3 \pm  0.5 $ & $55.6 \pm  1.2 $ & $32.8 \pm  0.9 $ & $  71 $& $ 204 $ \\ 
              DDO 53 &    3.7 &  50 &  51 &  90 & 15.48 & 15.48 & $11.0 \pm  0.4 $ & $17.8 \pm  0.4 $ & $31.1 \pm  0.8 $ &       $ <  0.4 $ & $ 0.8 $& $   3 $ \\ 
           M 81 Dw B &    8.9 &  29 &  55 & 110 & 14.47 & 15.02 & $ 5.0 \pm  0.4 $ & $11.0 \pm  0.4 $ & $14.7 \pm  1.3 $ &       $ <  0.8 $ & $ 0.8 $& -     \\ 
             NGC 784 &    5.2 &  80 & 210 &  90 & 12.14 & 12.14 & $51.9 \pm  4.6 $ & $79.7 \pm  4.0 $ & $125.5 \pm  6.1 $ &       $ <  4.2 $ & $   3 $& $ 116 $ \\ 
            NGC 3077 &    3.8 & 200 & 200 &  90 & 10.20 & 10.20 & $ 249 \pm   31 $ & $140.5 \pm 29.4 $ & $ 755 \pm   51 $ & $ 234 \pm   37 $ & $  29 $& $ 144 $ \\ 
            NGC 3274 &    6.8 & 120 &  40 &  90 & 12.98 & 13.03 & $32.7 \pm  1.5 $ & $76.1 \pm  1.6 $ & $95.4 \pm  2.5 $ & $ 8.9 \pm  1.7 $ & $   4 $& $  54 $ \\ 
            NGC 3738 &    5.3 & 100 &  60 &  90 & 12.18 & 12.18 & $51.5 \pm  1.5 $ & $103.7 \pm  1.5 $ & $ 145 \pm    2 $ & $13.1 \pm  2.0 $ & $   9 $& $  38 $ \\ 
            NGC 3741 &    3.2 &  55 &  60 &  90 & 14.13 & 14.31 & $12.7 \pm  0.7 $ & $22.9 \pm  0.7 $ & $38.4 \pm  1.3 $ &       $ <  0.6 $ & $   1 $& -     \\ 
            NGC 4068 &    4.4 & 110 & 120 &  90 & 13.21 & 13.21 & $17.9 \pm  1.7 $ & $35.3 \pm  1.7 $ & $55.0 \pm  3.3 $ &       $ <  1.8 $ & -    & -     \\ 
            NGC 4144 &    6.8 & 110 & 180 & 180 & 11.90 & 11.90 & $48.2 \pm  4.1 $ & $100.9 \pm  4.4 $ & $ 163 \pm    6 $ & $16.3 \pm  4.9 $ & $   8 $& $  14 $ \\ 
            NGC 4190 &    2.8 & 100 &  90 &  90 & 12.92 & 12.96 & $19.9 \pm  1.7 $ & $24.0 \pm  1.5 $ & $53.1 \pm  2.7 $ & $ 1.8 \pm  2.0 $ & $   5 $& $  62 $ \\ 
            NGC 4214 &    2.9 & 190 & 120 &  90 & 10.54 & 10.54 & $ 597 \pm    9 $ & $1441 \pm   10 $ & $1813 \pm   22 $ & $ 200 \pm   14 $ & $  38 $& $ 230 $\tablenotemark{a} \\ 
            NGC 4288 &    9.2 & 100 &  90 &  90 & 13.00 & 13.06 & $29.5 \pm  3.3 $ & $42.8 \pm  2.5 $ & $78.1 \pm  3.9 $ & $11.8 \pm  3.5 $ & $   7 $& $  78 $ \\ 
            NGC 4455 &    7.2 &  70 & 100 &  90 & 12.86 & 12.89 & $28.3 \pm  2.1 $ & $50.8 \pm  1.9 $ & $82.5 \pm  3.1 $ & $ 7.8 \pm  2.8 $ & -    & $  58 $ \\ 
            NGC 4605 &    5.5 & 100 & 230 &  25 & 10.55 & 10.56 & $ 192 \pm    8 $ & $ 232 \pm    8 $ & $ 654 \pm   17 $ & $ 127 \pm    9 $ & $  83 $& $ 173 $ \\ 
            NGC 4618 &    7.9 & 180 & 180 &  90 & 11.03 & 11.03 & $ 148 \pm    8 $ & $ 193 \pm    9 $ & $ 410 \pm   15 $ & $78.6 \pm 11.0 $ & $  36 $& $  95 $ \\ 
            NGC 4656 &    4.8 & 110 & 340 & 135 & 10.74 & 11.16 & $ 262 \pm    9 $ & $ 776 \pm    9 $ & $ 742 \pm   14 $ & $31.5 \pm 10.4 $ & $  60 $& -     \\ 
            NGC 4736 &    4.6 & 300 &  56 & 118 &  9.01 &  9.01 & $ 215 \pm   16 $ & $ 198 \pm   16 $ & $ 806 \pm   33 $ & $ 465 \pm   23 $ & $ 265 $& $ 792 $\tablenotemark{a} \\ 
            NGC 5238 &    4.5 &  85 &  50 &  90 & 13.35 & 14.02 & $20.4 \pm  0.8 $ & $48.0 \pm  0.8 $ & $54.5 \pm  1.7 $ & $ 3.6 \pm  1.2 $ & -    & $   4 $ \\ 
            NGC 5474 &    7.0 & 158 &  56 &  90 & 11.36 & 12.36 & $37.4 \pm  2.5 $ & $51.7 \pm  2.4 $ & $91.1 \pm  4.0 $ & $13.2 \pm  3.4 $ & $  12 $& -     \\ 
             UGC 685 &    4.7 &  80 &  40 &  90 & 13.96 & 14.17 & $ 9.5 \pm  0.8 $ & $17.5 \pm  0.7 $ & $25.6 \pm  1.4 $ & $ 1.8 \pm  1.1 $ & -    & -     \\ 
            UGC 4787 &    8.6 &  40 & 120 &  90 & 14.03 & 14.03 & $ 6.5 \pm  1.3 $ & $ 5.5 \pm  1.2 $ & $13.6 \pm  2.2 $ &       $ <  1.3 $ & -    & -     \\ 
            UGC 6541 &    4.2 &  65 &  30 &  90 & 14.04 & 14.36 & $21.2 \pm  0.7 $ & $90.2 \pm  0.7 $ & $56.0 \pm  1.0 $ &       $ <  0.6 $ & -    & $  19 $ \\ 
            UGC 7950 &    8.9 &  60 &  40 &  90 & 13.03 & 13.81 & $ 8.0 \pm  0.5 $ & $12.9 \pm  0.5 $ & $20.9 \pm  1.2 $ & $ 2.0 \pm  0.9 $ & $ 0.6 $& $   8 $ \\ 
            UGC 8508 &    2.6 &  90 &  60 &  90 & 13.68 & 13.92 & $ 8.8 \pm  1.0 $ & $ 7.7 \pm  1.0 $ & $26.0 \pm  2.3 $ &       $ <  0.8 $ & -    & -     \\ 
            UGCA 225 &   11.0 &  25 &  20 &  90 & 15.15 & 15.15 & $22.9 \pm  0.2 $ & $93.3 \pm  0.3 $ & $65.0 \pm  0.5 $ & $ 1.8 \pm  0.3 $ & $   2 $& -     \\ 
            UGCA 281 &    5.7 &  50 &  30 &  90 & 14.31 & 14.32 & $47.4 \pm  0.5 $ & $ 244 \pm    1 $ & $ 136 \pm    1 $ & $ 3.1 \pm  0.9 $ & $   4 $& $  38 $ \\ 

\\
\hline
\enddata
\tablenotetext{a}{${\rm 150~MHz}$ flux density from 6C or 7C \citep{6c1,6c2,6c3,6c4,6c5,6c6,7c}.}
\tablenotetext{b}{${\rm 150~MHz}$ flux density from GLEAM \citep{way15,hur16}.}
\end{deluxetable*}
\end{longrotatetable}

\clearpage

\begin{deluxetable*}{lll}
\tablecolumns{3}
\tabletypesize{\tiny}
\tablecaption{A selection of star formation rate indicator calibrations from the prior literature.\label{table:litcal}}
\tablehead{
  \colhead{Indicator\tablenotemark{a}} &
  \colhead{Fit\tablenotemark{b}} &
  \colhead{Reference} 
  }
\startdata
$ {\rm log} L_{FUV}\tablenotemark{c} $  & $ 42.03 + 0.74 \times ( {\rm log}  L_{H\alpha,{Corr}} - 40) $ & \citet{lee09} \\  
$ {\rm log} L_{FUV}\tablenotemark{c} $  & $ 42.09 + ( {\rm log}  L_{H\alpha,{Corr}} -40) $ & \citet{hao11} \\                     
$ {\rm log} L_{FUV}\tablenotemark{c} $  & $ 42.87 + 0.74 \times ( {\rm log}  L_{H\alpha,{Corr}} - 40) $ & \citet{dav16} \\ 
$ {\rm log} L_{FUV}\tablenotemark{c} $  & $ 41.70 + 1.11 \times ( {\rm log}  L_{H\alpha,{Corr}} - 40) $ & \citet{jai16} \\  
\\
$ {\rm log} L_{8~{\rm \mu m}} $  & $ 41.80 + 0.92 \times ( {\rm log}  L_{H\alpha,{Corr}} - 40) $ & \citet{wu05} \\                             
$ {\rm log} L_{8~{\rm \mu m}} $  & $ 41.56 + 0.94 \times ( {\rm log}  L_{H\alpha,{Corr}} - 40) $ & \citet{cal07}\tablenotemark{d}\tablenotemark{e} \\   
$ {\rm log} L_{8~{\rm \mu m}} $  & $ 41.97 + 1.14 \times ( {\rm log}  L_{H\alpha,{Corr}} - 40) $ & \citet{zhu08} \\                            
$ {\rm log} L_{8~{\rm \mu m}} $  & $ 41.67 + ( {\rm log}  L_{H\alpha,{Corr}} -40) $ & \citet{ken09} \\                                                
\\
$ {\rm log} L_{W3} $  & $ 41.61 + ( {\rm log}  L_{H\alpha,{Corr}} -40) $ & \citet{jar13} \\                                      
$ {\rm log} L_{W3} $  & $ 41.27 + 0.97 \times ( {\rm log}  L_{H\alpha,{Corr}} - 40) $ & \citet{lee13} \\                 
$ {\rm log} L_{W3} $  & $ 41.29 + 0.88 \times ( {\rm log}  L_{H\alpha,{Corr}} - 40) $ & \citet{clu14} \\                  
$ {\rm log} L_{W3} $  & $ 41.67 + 0.83 \times ( {\rm log}  L_{H\alpha,{Corr}} - 40) $ & \citet{dav16} \\                  
\\
$ {\rm log} L_{W4} $  & $ 41.43 + ( {\rm log}  L_{H\alpha,{Corr}} -40) $ & \citet{jar13} \\                                      
$ {\rm log} L_{W4} $  & $ 41.15 + 1.04 \times ( {\rm log}  L_{H\alpha,{Corr}} - 40) $ & \citet{lee13} \\                 
$ {\rm log} L_{W4} $  & $ 40.61 + 1.22 \times ( {\rm log}  L_{H\alpha,{Corr}} - 40) $ & \citet{clu14} \\                  
$ {\rm log} L_{W4} $  & $ 41.26 + ( {\rm log}  L_{H\alpha,{Corr}} -40) $ & \citet{cat15} \\                                     
$ {\rm log} L_{W4} $  & $ 40.84 + 1.36 \times ( {\rm log}  L_{H\alpha,{Corr}} - 40) $ & \citet{cat15} \\                  
$ {\rm log} L_{W4} $  & $ 41.33 + 1.20 \times ( {\rm log}  L_{H\alpha,{Corr}} - 40) $ & \citet{dav16} \\                  
\\
$ {\rm log} L_{24~{\rm \mu m}} $  & $ 41.11 + 1.12 \times ( {\rm log}  L_{H\alpha,{Corr}} - 40) $ & \citet{wu05} \\                               
$ {\rm log} L_{24~{\rm \mu m}} $  & $ 41.13 + 1.13 \times ( {\rm log}  L_{H\alpha,{Corr}} - 40) $ & \citet{cal07} \\                                       
$ {\rm log} L_{24~{\rm \mu m}} $  & $ 41.12 + 1.21 \times ( {\rm log}  L_{H\alpha,{Corr}} - 40) $ & \citet{rel07} \\                               
$ {\rm log} L_{24~{\rm \mu m}} $  & $ 41.10 + 1.18 \times ( {\rm log}  L_{H\alpha,{Corr}} - 40) $ & \citet{zhu08} \\                                
$ {\rm log} L_{24~{\rm \mu m}} $  & $ 41.33 + ( {\rm log}  L_{H\alpha,{Corr}} -40) $ & \citet{ken09} \\                                                 
$ {\rm log} L_{24~{\rm \mu m}} $  & $ 41.53 + 1.18 \times ( {\rm log}  L_{H\alpha,{Corr}} - 40) $ & \citet{rie09}\tablenotemark{e}  \\   
\\
$ {\rm log} P_{1.4~GHz} $  & $ 20.20 + ( {\rm log} L_{H\alpha,{Corr}} -40) $ & \citet{con92} \\                              
$ {\rm log} P_{1.4~GHz} $  & $ 20.16 + {\rm log}  ( L_{H\alpha,{Corr}} -40) $ when $ {\rm log} P_{1.4 GHz} > 21.81 $ & \citet{bel03} \\  
$ {\rm log} P_{1.4~GHz} $  & $ 20.05 + {\rm log}  ( L_{H\alpha,{Corr}} -40) $ & \citet{ken09} \\                             
$ {\rm log} P_{1.4~GHz} $  & $ 19.62 + 1.18 \times ( {\rm log}  L_{H\alpha,{Corr}} - 40) $ & \citet{bos15} \\           
\hline
\enddata
\tablenotetext{a}{UV and mid-infrared luminosities are presented in units of ${\rm erg~s^{-1}}$ while radio powers are presented in units of ${\rm W~Hz^{-1}}$.}
\tablenotetext{b}{In some instances we have converted SFRs to $L_{H\alpha,{Corr}}$ using
$SFR({\rm M_\odot~ yr^{-1}}) =  7.9\times 10^{-42} L_{\rm H\alpha} ({\rm erg~s^{-1}})$ for a \citet{sal55} IMF,
$SFR({\rm M_\odot~ yr^{-1}}) =  5.5\times 10^{-42} L_{\rm H\alpha} ({\rm erg~s^{-1}})$ for a \citet{kro01} IMF, 
$SFR({\rm M_\odot~ yr^{-1}}) =  1.2\times 10^{-41} L_{\rm H\alpha} ({\rm erg~s^{-1}})$ for a \citet{cha03} IMF, and 
$SFR({\rm M_\odot~ yr^{-1}}) =  5.1\times 10^{-42} L_{\rm H\alpha} ({\rm erg~s^{-1}})$ for a \citet{bal03} IMF.}
\tablenotetext{c}{GALEX $FUV$ luminosities have been corrected for dust extinction, and we refer readers to the original papers for relevant details.}
\tablenotetext{d}{The \citet{cal07} $8~{\rm \mu m}$ relation is for luminosity per ${\rm kpc^2}$ }
\tablenotetext{e}{We adopt $L_{\rm Pa\alpha} = 0.128 L_{\rm H\alpha}$ \citep{hum87}.}
\end{deluxetable*}

\clearpage

\begin{rotatetable}
\begin{deluxetable*}{llcccc}
\tablecolumns{6}
\tabletypesize{\tiny}
\tablecaption{Star formation rate indicator calibrations.\label{table:calibration}}
\tablehead{
  \colhead{Indicator\tablenotemark{a}} &
  \colhead{Fit} &
  \colhead{$\sigma_{H\alpha,BPT}$} &
  \colhead{$\sigma_{H\alpha,More}$\tablenotemark{b}} &
  \colhead{$>2\sigma$} &
  \colhead{$n$}\\
  \colhead{} &
  \colhead{} &
  \colhead{(dex)} &
  \colhead{(dex)} &
  \colhead{fraction} &
  \colhead{}  
}
\startdata
$ {\rm log} L_{FUV} + 2 \times (M_{FUV}-M_{NUV})  $  & $(42.42 \pm 0.05) + (0.96 \pm 0.03) \times ( {\rm log}  L_{H\alpha,{Corr}} - 40)  $ & 0.35 & 0.39 & 0.03 & 62 \\
$ {\rm log} L_{FUV} + 1.532 \times (M_{FUV}-M_{NUV})-0.0088 $  & $(42.25 \pm 0.04) + (0.90 \pm 0.03) \times ( {\rm log}  L_{H\alpha,{Corr}} - 40)  $ & 0.29 & 0.29 & 0.06 & 62 \\
$ {\rm log} L_{8~{\rm \mu m}}  $  & $(40.88 \pm 0.07) + (1.30 \pm 0.05) \times ( {\rm log}  L_{H\alpha,{Corr}} - 40)  $ & 0.33 & 0.37 & 0.07 & 60 \\ 
$ {\rm log} L_{W3}  $  & $(40.79 \pm 0.06) + (1.27 \pm 0.04) \times ( {\rm log}  L_{H\alpha,{Corr}} - 40)  $ & 0.28 & 0.34 & 0.05 & 61 \\ 
$ {\rm log} L_{W4}  $  & $(40.96 \pm 0.04) + (1.26 \pm 0.03) \times ( {\rm log}  L_{H\alpha,{Corr}} - 40)  $ & 0.20 & 0.27 & 0.05 & 58 \\ 
$ {\rm log} L_{24~{\rm \mu m}}  $  & $(40.93 \pm 0.04) + (1.30 \pm 0.03) \times ( {\rm log}  L_{H\alpha,{Corr}} -40)  $ & 0.18 & 0.24 & 0.08 & 62 \\ 
$ {\rm log} P_{\rm 1.4~GHz}  $  & $(19.65 \pm 0.05) + (1.27 \pm 0.03) \times ( {\rm log}  L_{H\alpha,{Corr}} - 40)  $ & 0.18 & 0.22 & 0.08 & 52 \\ 
$ {\rm log} P_{\rm 150~MHz}  $  & $(20.49 \pm 0.08) + (1.16 \pm 0.05) \times ( {\rm log}  L_{H\alpha,{Corr}} - 40)  $ & 0.24 & 0.32 & 0.08 & 36 \\ 
\\
$ {\rm log} L_{8~{\rm \mu m}}  $  & $ (40.49 \pm 0.08) + ( {\rm log}  L_{H\alpha,{Corr}} - 40) + (0.38 \pm 0.04) \times ({\rm log} L_{4.5~{\rm \mu m}} - 40) $ & 0.35 & 0.36 & 0.05 & 60 \\ 
$ {\rm log} L_{W3}  $  & $ (40.52 \pm 0.05) + ( {\rm log}  L_{H\alpha,{Corr}} - 40) + (0.31 \pm 0.03) \times ({\rm log} L_{W2} - 40) $ & 0.25 & 0.29 & 0.05 & 61 \\ 
$ {\rm log} L_{W4}  $  & $ (40.79 \pm 0.05) + ( {\rm log}  L_{H\alpha,{Corr}} - 40) + (0.25 \pm 0.02) \times ({\rm log} L_{W2} - 40) $ & 0.23 & 0.30 & 0.03 & 58 \\ 
$ {\rm log} L_{24~ {\rm \mu m}}  $  & $ (40.69 \pm 0.05) + ( {\rm log}  L_{H\alpha,{Corr}} - 40) + (0.29 \pm 0.03) \times ({\rm log} L_{4.5 {\rm \mu m}} - 40) $ & 0.26 & 0.30 & 0.02 & 62 \\ 
$ {\rm log} P_{\rm 1.4~GHz}  $  & $ (19.65 \pm 0.05) + ( {\rm log}  L_{H\alpha,{Corr}} - 40) + (0.27 \pm 0.03) \times ({\rm log} L_{W2} - 40) $ & 0.22 & 0.27 & 0.08 & 52 \\ 
$ {\rm log} P_{\rm 150~MHz}  $  & $ (20.49 \pm 0.08) + ( {\rm log}  L_{H\alpha,{Corr}} - 40) + (0.16 \pm 0.05) \times ({\rm log} L_{W2} - 40) $ & 0.28 & 0.37 & 0.08 & 36 \\ 
\hline
\enddata
\tablenotetext{a}{UV and mid-infrared luminosities are presented in units of ${\rm erg~s^{-1}}$ while radio powers are presented in units of ${\rm W~Hz^{-1}}$.}
\tablenotetext{b}{$\sigma_{H\alpha,More}$ is measured using galaxies that meet the less conservative BPT criterion of \citet{kew01}, which may include some AGNs that inflate the scatter.}
\end{deluxetable*}
\end{rotatetable}

\clearpage


\bibliographystyle{astroads}
\bibliography{ms.bib}

\end{document}